\documentclass[aps,prd,twocolumn,superscriptaddress,amssymb,floatfix,nobibnotes,nofootinbib,a4paper,noraggedfooter,flushbottom,showpacs]{revtex4}
\usepackage{epsf}
\usepackage{graphicx}
\usepackage{amsmath}
\begin{document}
\newcommand{\tr}{\mbox{tr}\,}
\newcommand{\Tr}{\mbox{Tr}\,}
\newcommand{\www}{\mbox{
\begin{minipage}{16pt}\includegraphics[width=16pt]{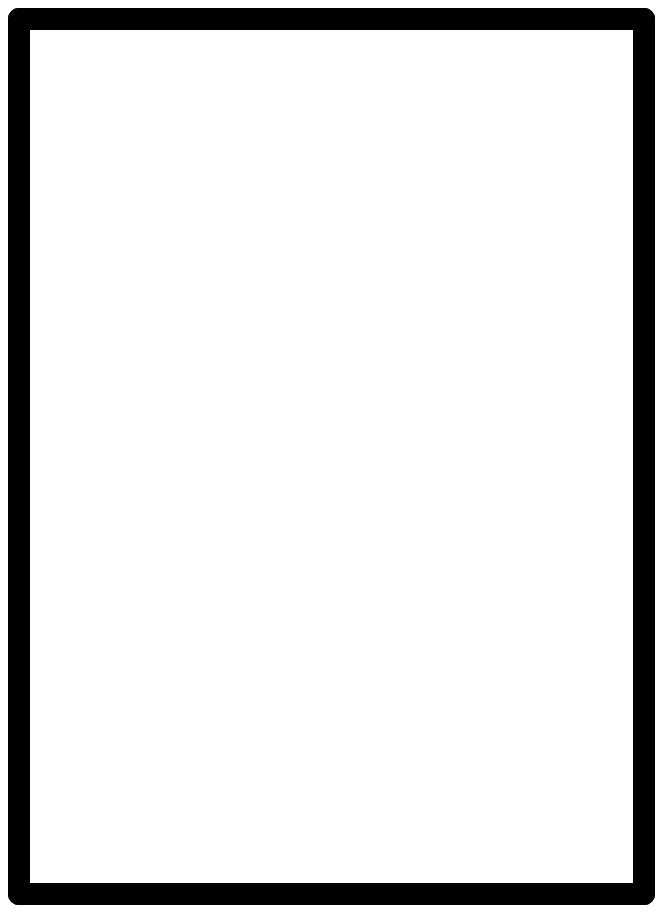}\end{minipage}}}
\newcommand{\wbbc}{\mbox{
\begin{minipage}{16pt}\includegraphics[width=16pt]{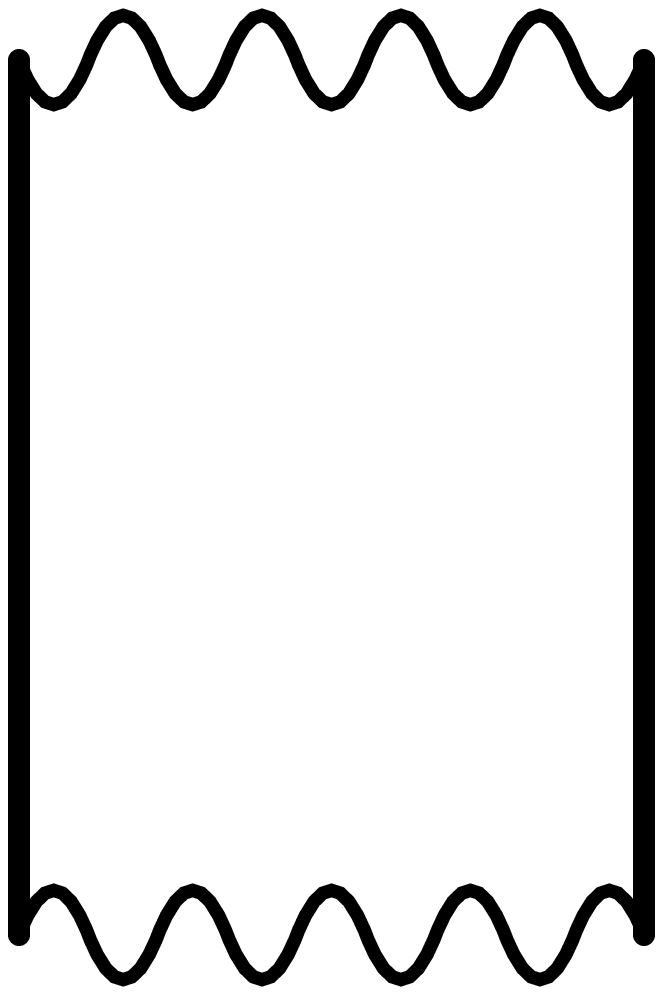}\end{minipage}}}
\newcommand{\wbbd}{\mbox{
\begin{minipage}{16pt}\includegraphics[width=16pt]{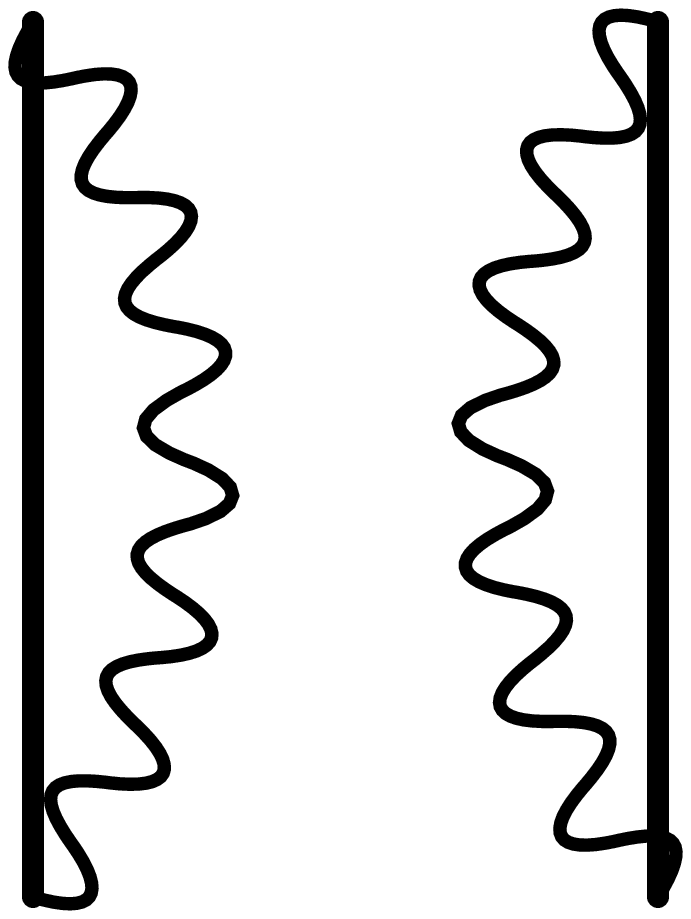}\end{minipage}}}
\newcommand{\wwb}{\mbox{
\begin{minipage}{16pt}\includegraphics[width=16pt]{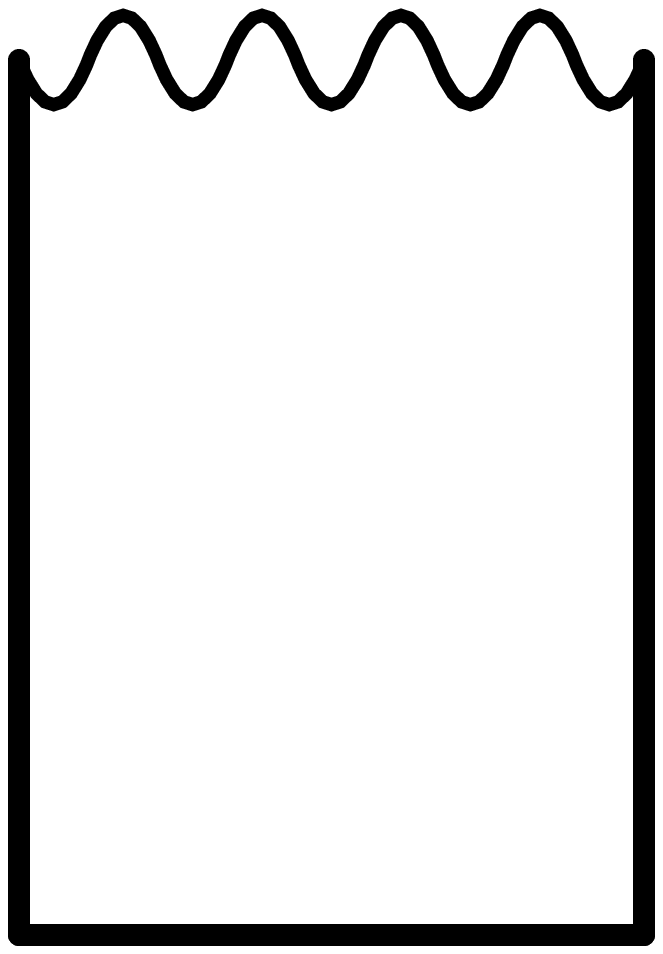}\end{minipage}}}
\newcommand{\wbw}{\mbox{
\begin{minipage}{16pt}\includegraphics[width=16pt]{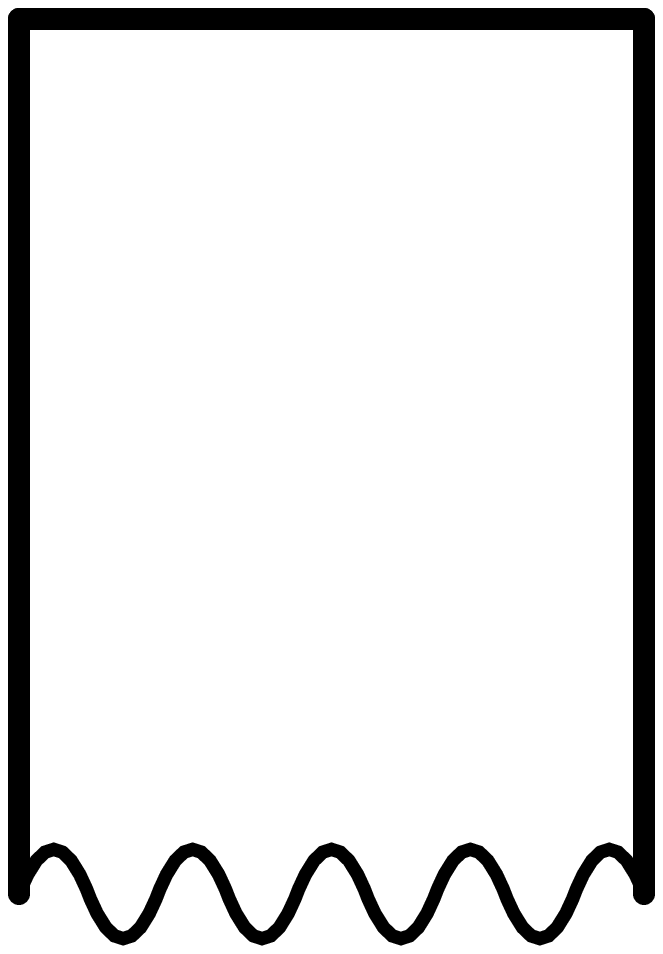}\end{minipage}}}
\preprint{???}
\title
{Observation of String Breaking in QCD}
\author{Gunnar S.\ Bali}
\email{g.bali@physics.gla.ac.uk}
\affiliation{Department of Physics \& Astronomy, The University of Glasgow, Glasgow G12 8QQ, Scotland}
\author{Hartmut Neff}
\email{hneff@buphy.bu.edu}
\affiliation{Center for Computational Science, Boston University, 3 Cummington
St, Boston MA02215, USA}
\author{Thomas D\"ussel}
\email{th.duessel@fz-juelich.de}
\author{Thomas Lippert}
\email{th.lippert@fz-juelich.de}
\affiliation{John von Neumann Institute for Computing,
Forschungszentrum J\"ulich, D-52425 J\"ulich, Germany}
\author{Klaus Schilling}
\email{schillin@theorie.physik.uni-wuppertal.de}
\affiliation{Fachbereich Physik, Bergische Universit\"at Wuppertal,
Gau\ss{}stra\ss{}e, D-42097 Wuppertal, Germany}
\collaboration{SESAM Collaboration}
\date{\today}
\begin{abstract}
We numerically investigate the transition
of the static quark-antiquark string into a static-light
meson-antimeson system.
Improving noise reduction techniques, 
we are able to resolve
the signature of string breaking dynamics 
for $n_f=2$ lattice QCD at zero temperature.
This result can be related to
properties of quarkonium systems. We also study short-distance
interactions between
two static-light mesons.
\end{abstract}
\maketitle

\section{Introduction}

Sea quarks are an important ingredient of strong interaction dynamics. In the
framework of quantum chromodynamics, however, quantitative calculations of
their effects on hadron phenomenology have proven
to be notoriously difficult, unless
one resorts to approximations based on additional model assumptions.
Nevertheless, the {\it ab initio} approach of lattice gauge theory
towards the sea quark problem has shown steady progress over the past decade:
recently the $\eta '$-problem has been tackled successfully  on the
lattice~\cite{Schilling:2004kg,Struckmann:2000bt,McNeile:2001cr} where sea
quarks induce the axial anomaly in the sense of the Witten-Veneziano
mechanism~\cite{Witten:1979vv,Veneziano:1980xs}.

Another example is the strong decay of hadrons
through light quark-antiquark pair creation, for instance
the transition from a colour string configuration between two
static colour sources, $\overline{Q}Q$, into a pair of static-light mesons,
$B\overline{B}$.  This colour string breaking, which we address in
this paper, is expected to occur as soon
as the colour source-sink separation, $r$, exceeds a certain threshold value,
$r_c>1\,\mbox{fm}$.

In lattice simulations this behaviour has been investigated
in four dimensional QCD at zero temperature $T$ with sea
quarks~\cite{Bali:2000vr,bolder,Aoki:1998sb,
  Bonnet:1999gt,Pennanen:2000yk,Duncan:2000kr,Bernard:2001tz} as well as in
QCD${}_3$~\cite{Trottier:2004qc}. However, these studies lacked
compelling evidence of string breaking\footnote{The $T>
  0$ situation appears to be more favourable \cite{Laermann:1998gm}.}.
This failure is due to problems like:
{\em (i)}
String breaking investigations only make sense in a full QCD
setting with large ensemble sizes.
{\em (ii)}
String breaking occurs at distances beyond 1~fm, a regime with
a poor signal-to-noise ratio.
{\em (iii)}
The poor overlap of the $\overline{Q}Q$ creation operator with
the large-distance $B\overline{B}$ ground state.

This last problem necessitates to resolve the signal
at huge Euclidean times $t$,
unless one bases the investigation on a $2\times 2$
correlation matrix, whose additional elements include the
insertion of light quark
propagators into the standard Wilson
loop~\cite{Pennanen:2000yk,Duncan:2000kr,Bernard:2001tz,bolder}. Such
quark insertions require propagators from any source to any sink
position (``all-to-all propagators''), in order to enable the
exploitation of translational invariance
for error reduction (self averaging).

For QCD with $n_f$ mass-degenerate sea quark flavours this correlation
matrix takes the form,
\begin{eqnarray}
\nonumber
C(t)\quad=&&\left(\begin{array}{cc}C_{QQ}(t)&C_{QB}(t)\\
C_{BQ}(t)&C_{BB}(t)\end{array}\right)\\
=e^{-2m_Qt}\!\!\!\!\!&&\left(\begin{array}{rl}
\quad \www&\quad \sqrt{n_f}\wwb\\&\\
\sqrt{n_f}\wbw&\quad -n_f\wbbc+\wbbd\end{array}\right),\label{eq:sb}
\end{eqnarray}
where the straight lines denote gauge transporters and the wiggly lines
represent light quark
propagators\footnote{Details of this expression will be discussed in
  Sec.~\ref{sec:mixing} below.}. We refer to the difference between
the physical eigenstates and the $\overline{Q}Q$ and $B\overline{B}$ basis
as ``mixing''. Such mixing should manifest itself
``explicitly'', by non-vanishing off-diagonal matrix elements, relative
to the diagonal matrix elements, and
``implicitly''. The latter refers either
to the Wilson loop
$C_{QQ}(t)$ decaying into the mass of the (dominantly) $B\overline{B}$ state
for $r>r_c$ or to a decay of $C_{BB}(t)$ towards the
$\overline{Q}Q$ mass for $r<r_c$, as $t\rightarrow\infty$.
Implicit mixing is much harder to detect than explicit mixing.

In the quenched approximation baryon and anti-baryon numbers are separately
conserved and the $\overline{Q}Q$ and $B\overline{B}$ sectors are mutually
orthogonal. By definition, $n_f$ vanishes and there will be no mixing.
This does not mean however, that the matrix elements
accompanying the explicit
$n_f$- and $\sqrt{n_f}$-factors are necessarily {\em zero}.

So far string breaking has been verified in the following cases:
in $SU(2)$ gauge theory with a fundamental scalar field in three
dimensions~\cite{Philipsen:1998de} and in four
dimensions~\cite{Jersak:1988bf,Bock:1988kq,Knechtli:1998gf} as well as for the
$SU(2)$ potential between adjoint sources (screened by the gluons) in three
dimensions~\cite{Poulis:1995nn,Stephenson:1999kh,Philipsen:1999wf,Kratochvila:2003zj}
and in four dimensions~\cite{deForcrand:1999kr,Kallio:2000jc}.
However, in only one of these studies~\cite{Kratochvila:2003zj},
and in a recent simulation of the 3d ${\mathbb Z}_2$-Higgs
model~\cite{Gliozzi:2004cs},
{\em implicit} string breaking has been convincingly demonstrated.

Let us recall the signature of string breaking: without mixing,
the $\overline{Q}Q$ and the $B\overline{B}$ are QCD eigenstates
and will undergo a
plain level crossing (with minimal energy gap $\Delta E_c=0$),
at a certain critical
distance $r_c$.  Simulations with $n_f=0$ show exactly this
behaviour.
In contrast, with $q\bar{q}$ creation/annihilation switched on,
the Fock states will
undergo sizeable mixing in the neighbourhood of $r_c$.
The minimal energy gap $\Delta E_c$
between the two eigenstates
will grow with the spatial width of the mixing region.

The transition rate between $\overline{Q}Q$ and $B\overline{B}$
states is given by the (normalized)
time derivative of the off-diagonal matrix element,
$g=[dC_{QB}(t)/dt]_{t=0}[C_{BB}(0)C_{QQ}(0)]^{-1/2}$.
{}From a string picture as well as from strong coupling
arguments one would 
expect a more pronounced mixing in larger space-time dimensions $d$.
Therefore, the size of the energy gap within the string breaking
region should increase as one goes from $d=3$ to $d=4$.
In the large $N_c$ limit
we find, $g\propto \sqrt{n_f/N_c}$, for the potential between fundamental
sources, screened by $n_f$ flavours of fundamental scalar or quark fields.
For the breaking of the adjoint potential this translates into,
$g\propto 1/N_c$.

The
figures of Refs.~\cite{Philipsen:1999wf,Stephenson:1999kh}
for the adjoint string in 3d
$SU(2)$ gauge theory suggest the following upper limits for the size
of the energy gap, expressed in units of the string breaking distance:
$\Delta E_c r_c<0.45$ and $\Delta E_c r_c<0.75$,  respectively.
In 4d $SU(2)$ gauge theory one finds,
$\Delta E_cr_c<0.6$~\cite{deForcrand:1999kr}, whereas
Refs.~\cite{Philipsen:1998de,Knechtli:1998gf} show that the fundamental
3d $SU(2)$ string, screened by a scalar
field, satisfies $\Delta E_cr_c<0.35$ and
$\Delta E_c r_c<0.65$,
respectively. 
In all these cases either the spatial resolution of the string breaking region
was too coarse or the statistical errors were too large
to allow for the determination of
a lower bound.

Based on the qualitative
$n_f$, $N_c$ and $d$-dependencies discussed above, we expect the $n_f=2$
QCD energy gap to be somewhat bigger than the gaps quoted for the
toy model studies. From this reasoning we would aim at an error
of $\Delta E_cr_c$, smaller than 0.1. To meet this constraint,
we require a distance resolution in the string breaking region of
$\Delta r<0.1/(\sigma r_c) \approx 0.02\, r_c\approx 0.025$~fm,
where $\sigma$ denotes the
string tension.
We will find, $\Delta E_c r_c= 0.33(5)$.

In order to achieve the required precision, we apply a fourfold arsenal
of critical improvements, within
the $2\times 2$ correlation matrix setting:
\begin{description}
\item[Ground state overlaps.] It is essential to achieve a large overlap between the
  trial wavefunctions and the respective physical ground states.
 This
 enhances the signal since it will decay less rapidly with Euclidean time.
  Moreover, the large $t$ asymptotics will be reached at smaller temporal
  distances, further reducing the noise/signal ratio.  To this end we
  employ combinations of APE and Wuppertal smearing techniques
(see Sec.~\ref{sec:over}).
\item[Wilson loops.] The Wilson loop signal $C_{QQ}(t)$ can be further
  enhanced by using an improved fat link static action.  In this way, the
  relative errors of the Wilson loop data
are reduced by factors of about {\em five}
  (see Sec.~\ref{sec:over}).
\item[Quark propagators.] The generalized Wilson
  loops of Eq.~(\ref{eq:sb}) require the computation
of all-to-all light quark
  propagators if we wish to
fully exploit self averaging.
 Direct inversion of the Wilson
  Dirac matrix $M$ would be computationally prohibitive. Therefore, we
  approximate $M^{-1}$ by the lowest lying eigenvectors
  of $\gamma_5M$ using the truncated eigenmode approach (TEA)
\cite{Neff:2001zr},
together with a stochastic estimation
(SET, see e.g.\ the review~\cite{Wilcox:1999ab}) in the orthogonal subspace.
  Moreover, we apply a
  ``hopping parameter acceleration'' (HPA) for variance reduction. This further
reduces the errors
 of the
  disconnected contribution to $C_{BB}$ by
  factors of about {\em three} (see Sec.~\ref{sec:alltoall}).
\item[Distance resolution.] In order to avoid
  finite size effects and to achieve a fine distance resolution of the string
  breaking region, we employ a large set of
 off-axis distances (see Sec.~\ref{sec:simul}). 
\end{description}

With these methods we are able to demonstrate
compelling evidence, both for explicit mixing and for
string breaking in full QCD,
as well as for implicit mixing within $C_{BB}(t)$ for $r<r_c$.
We find the breaking of the quark antiquark string to occur
at a distance $r_c\approx
15\,a\approx 2.5\,r_0\approx 1.25$~fm, in units of $r_0\approx
0.5$~fm~\cite{Sommer:1993ce,Bali:1998pi}.

Note that our study should be viewed as exploratory since we restrict
ourselves to {\em
  one} value of the sea quark mass (slightly below the physical strange quark
mass), at {\em one} lattice spacing.  For details on our simulation parameters,
see Sec.~\ref{sec:simul}.

The paper is organized as follows: in Sec.~\ref{sec:mixing} we discuss the
mixing problem in detail. The notation used within Eq.~(\ref{eq:sb}) will be
defined.  We describe the combined application of TEA, SET and HPA for
the calculation of all-to-all propagators in Sec.~\ref{sec:latt}.  In
Sec.~\ref{sec:simu} we discuss theoretical expectations for the individual
matrix elements and check these against our data.  In
Sec.~\ref{sec:result}, we present and interpret
our main result, string breaking
in QCD, as well as the short-distance interactions between two static-light
mesons with isospin $I=0$ and $I=1$. We comment on the
phenomenological implications in Sec.~\ref{sec:pheno}.

In view of the length of this paper, we kept the sections as self-contained as
possible. For instance, the reader who is less interested in the technical
aspects of the study can safely skip
Sec.~\ref{sec:latt} altogether and concentrate on Secs.~\ref{sec:result}
and \ref{sec:pheno}.

\section{The mixing problem}
\label{sec:mixing}
Let us consider a system with a heavy quark $Q$ and a heavy
antiquark $\overline{Q}$ in
the static approximation, with separation $r=|{\mathbf R}|/a$, where $a$
denotes the lattice spacing and ${\mathbf R}$ is an integer valued
three-vector. We restrict our discussion to the $\Sigma_g^+$
ground state of the static system, with cylindrical symmetry.

Without sea quarks, the energy of this static-static system will
linearly diverge with
$r$~\cite{Bali:1992ru,Booth:1992bm,Bali:2000gf} as $r\rightarrow\infty$.  In
the presence of sea quarks, however, there will be some critical ``string
breaking'' distance $r_c$: when $r$ exceeds $r_c$, the mass of a system
containing two static-light mesons, which we shall call $B$ and
$\overline{B}$, separated by $r$, will become energetically favoured.  The
static $\overline{Q}Q$ potential will exhibit screening and saturate towards
about twice the mass of the $B$ meson.

A full investigation of this phenomenon requires the study of the Green
functions that correspond to the propagation of the $\overline{Q}Q$ and
$B\overline{B}$ systems as well as of the transition element between
these two states.
We start by defining our notations and discussing the symmetries
of the problem,
before we display the relevant Green functions.

\subsection{Definitions and representations}

The Euclidean Dirac equation in the static limit,
\begin{equation}
\label{eq:static}
\left(D_4\gamma_4+m_Q\right)|Q\rangle=0,
\end{equation}
yields the propagators of static quark and antiquark:
\begin{eqnarray}\label{eq:quark}
Q_y\overline{Q}_x&=&
\langle y|Q\rangle\langle Q|\gamma_4|x\rangle\\\nonumber&=&\delta_{{\mathbf x}{\mathbf y}}
U_{{\mathbf x}}(y_4,x_4)e^{-m_Q(y_4-x_4)}P_+,\\
\overline{Q}^{\dagger}_yQ^{\dagger}_x&=&
\langle y|\overline{Q}\rangle\langle \overline{Q}|\gamma_4|x\rangle
\\\nonumber&=&
\delta_{{\mathbf x}{\mathbf y}}
U^{\dagger}_{{\mathbf x}}(y_4,x_4)e^{-m_Q(y_4-x_4)}P_-,\label{eq:aquark}
\end{eqnarray}
where $y_4\geq x_4$ and,
\begin{equation}
P_{\pm}=\frac{1\pm\gamma_4}{2},
\end{equation}
are projectors onto the upper and lower two Dirac components.  $U_{{\mathbf
    x}}(y_4,x_4)\in SU(3)$ denotes a lattice discretization of the Schwinger
line, connecting $({\mathbf x},x_4)$ with $({\mathbf x},y_4)$:

\begin{equation}
\label{eq:schwinger}
U_{{\mathbf x}}(y_4,x_4)\simeq T\,\exp\left\{
ig\int_{x_4}^{y_4}\!\!\!\!dt\,A_4({\mathbf x},t)\right\}.
\end{equation}
$T$ denotes the time ordering operator. We use the convention $\langle
x|Q\rangle=Q_x$ and $\langle Q|x\rangle=Q^{\dagger}_x$, {\em i.e.}  \ 
$\overline{Q}_x=\langle Q|\gamma_4|x\rangle$.  

$m_Q(a)$ in Eqs.~(\ref{eq:quark}) and (\ref{eq:aquark}) above is the heavy
quark mass in a lattice scheme\footnote{$m_Q$ contains a power term in the
  inverse lattice spacing $a^{-1}$, $\delta m\propto \alpha_L/a +\cdots$,
  where $\alpha_L=g^2/(4\pi)$ is the strong coupling parameter in the lattice
  scheme. This term (that diverges in the continuum limit $a\rightarrow 0$)
  cancels against a similar contribution from $U_{{\mathbf x}}(t_2,t_1)$.
We shall also refer to this contribution as the ``self-energy'' associated
with the static propagator. Note that factorising $m_Q(a)$ into pole mass and
self-energy introduces a renormalon ambiguity, see
e.g.~\cite{Martinelli:1995vj,Bali:2003jq}.}.
We define the light quark Dirac operator,
\begin{equation}
M=\left(|q\rangle\langle q|\gamma_4\right)^{-1}=D_{\mu}\gamma_\mu+m.
\end{equation}
We use Wilson fermions throughout the paper:
\begin{eqnarray}
\nonumber
M_{xy}
=\delta_{xy}-\kappa\sum_{\mu=1}^4&&\left[(1-\gamma_{\mu})U_{x,\mu}\delta_{x+a\hat{\mu},y}\right.\\\label{eq:wilferm}
&&+\left.(1+\gamma_{\mu})U_{x,-\mu}\delta_{x-a\hat{\mu},y}\right].
\end{eqnarray}
As usual, the quark fields have been rescaled by factors $\sqrt{2\kappa}$
where $\kappa= (8+2ma)^{-1}$ in the free field case and in general,
$\kappa(m=0)=\kappa_c\geq 1/8$. $U_{x,\mu}$ denotes an $SU(3)$ gauge field and 
$U_{x,-\mu}=U^{\dagger}_{x-a\hat{\mu},\mu}$, 
$M^{\dagger}=\gamma_5M\gamma_5$.

We define a gauge transporter $V_t({\mathbf y},{\mathbf x})\in SU(3)$,
connecting the point $({\mathbf x},t)$ with $({\mathbf y},t)$.  This is to be
taken local in time and rotationally symmetric about the shortest connection.
The properties under local gauge transformations $\Omega_x\in SU(3)$ are:
\begin{eqnarray}
Q_x\mapsto \Omega_xQ_x,\quad
\overline{Q}_x\mapsto \overline{Q}_x\Omega_x^{\dagger},\\
V_t({\mathbf y},{\mathbf x})\mapsto \Omega_{({\mathbf y},t)}
V_t({\mathbf y},{\mathbf x})\Omega^{\dagger}_{({\mathbf x},t)},
\end{eqnarray}
which implies that the combination,
\begin{equation}
\label{eq:gamma}
{\mathcal Q}_t({\mathbf y},{\mathbf x})=\overline{Q}_{({\mathbf y},t)}
\frac{\boldsymbol{\gamma}\cdot\mathbf{r}}{r}
V_t({\mathbf y},{\mathbf x})Q_{({\mathbf x},t)},
\end{equation}
is a colour singlet.

The spins of $\overline{Q}$ and $Q$ can either couple symmetrically or
anti-symmetrically. The first situation is represented by
${\boldsymbol{\gamma}\cdot\mathbf{r}}/{r}$ in Eq.~(\ref{eq:gamma}) (total
spin $S=1$), the latter choice corresponds to the replacement,
$\boldsymbol{\gamma}\cdot{\mathbf r}/r\mapsto\gamma_5$, within
Eq.~(\ref{eq:gamma}) ($S=0$), see also
Ref.~\cite{Drummond:1998ir}.

The relevant symmetry group is not $O(3)\otimes{\mathcal C}$
but its cylindrical subgroup $D_{\infty h}$. On the lattice this reduces to
$D_{4h}$. Nevertheless, we will use the continuum expressions,
as the ``latticization''
is straight forward in this case~\cite{Campbell:1987nv,Juge:1999ie}.  The
irreducible representations of $D_{\infty h}$ are conventionally labelled by
the spin along the axis, $\Lambda$, where $\Sigma,\Pi,\Delta$ refer to
$\Lambda=0,1,2$, respectively, with a subscript $\eta=g$ for
$\mathcal{CP}=+$ (gerade, even)
or $\eta=u$ for $\mathcal{CP}=-$ (ungerade, odd) 
transformation properties.  Parity $\mathcal P$ or charge $\mathcal C$ alone
are not ``good'' quantum numbers.  The $\Sigma$ representations carry, in
addition to the $\eta$ quantum number, an ${\mathcal R}$ parity with respect
to reflections on a plane that includes the two endpoints. This results
in an additional $\pm$ superscript.  The symmetric spin combination
Eq.~(\ref{eq:gamma}), when combined with a symmetric gluonic string $V_t$,
lies within the $\Sigma_g^+$ ground state representation while the
antisymmetric spin-combination, $\boldsymbol{\gamma}\cdot{\mathbf
  r}/r\mapsto\gamma_5$ corresponds to $\Sigma_u^-$. These two representations
yield degenerate energy levels, since both are calculated from one and the same
Wilson loop\footnote{The labelling is somewhat different if we start from
  scalar rather than from fermionic static sources. In this case the
  ground state $\Sigma_g^+$ potential is not accompanied by any other
  mass-degenerate states and $\Sigma_u^-$ would label a non-trivial
  gluonic excitation.}.

Once mass corrections are added to the static limit, the full
$O(3)\otimes{\mathcal C}$ symmetry becomes restored. The $\Sigma_g^+$
representation is contained within the $J^{PC}=0^{++}, 1^{--}, 2^{++},\cdots$
sectors of this bigger symmetry group while $\Sigma_u^-$ corresponds to
$J^{PC}=0^{-+}, 1^{+-},\cdots$. Within the two-quark sector, the $1^{--}$ and
$0^{-+}$ states form the respective (mass-degenerate)
ground states since the other quantum numbers
require angular momentum $L>0$ or non-trivial gluonic excitations.

\subsection{The elements of the correlation matrix}
We consider $n_f$ mass-degenerate flavours of light quarks $q^i$, $i=1,\ldots,
n_f$. Let $B_i$ be the $B$-meson with light quark flavour $i$.
For simplicity we
label the $\overline{Q}_{{\mathbf r}}Q_{{\mathbf 0}}$ string
creation operator as ${\mathcal Q}$
and the $B_{i,{\mathbf
    r}}\overline{B}_{i,{\mathbf 0}}$ operators as ${\mathcal B}_i
={\mathcal B}_{ii}$. We
suppress the distance ${\mathbf r}$ for ease of notation. 
We define the (unnormalized) states,
\begin{equation}
\label{eq:norm}
|Q\rangle={\mathcal Q}|0\rangle,\quad
|B_i\rangle={\mathcal B}_i|0\rangle.
\end{equation}
In what follows, $|Q\rangle$ will always denote this
$\overline{Q}Q$ state and should not be confused
with the static quark
spinor of the same name 
within Eqs.~(\ref{eq:static}) -- (\ref{eq:aquark}).
The lightest
static-light meson has light quark $J^P=\frac{1}{2}^-$,
see e.g.~\cite{Bali:2003jv} and references therein.  Combining this with
the heavy quark spin leads to mass-degenerate pseudoscalar and vector states.
Two of these pseudoscalars/vectors combined have $\mathcal{CP}=+$ and fall
into the $\Sigma_g^+$ representation (in
the vector case the spins have to be anti-aligned accordingly, to yield
$J_z=0$).  For concreteness we shall
choose\footnote{One can also work out the correlation
 matrix elements,
  starting from two vector states. Another possibility would be to probe the
  $\Sigma_u^-$ sector with an antisymmetric combination of vector and
  pseudoscalar $B$ states. As it should be, all these starting points yield
  identical Green functions (with the exception of $r=0$),
in the infinite quark mass
  limit.},  $B_i=\overline{Q}\gamma_5q^i$.
We are now in the position to display the three Green
functions that are relevant to our problem.  For the time evolution of
the $\overline{Q}Q$
state this reads:
\begin{widetext}
\begin{eqnarray}\nonumber
\left\langle Q|{\mathcal T}^{\tau}|Q\right\rangle_U&=&
\left\langle 0\left|
T\left\{\left[\overline{Q}_{({\mathbf r},t)}
\frac{\boldsymbol{\gamma}\cdot\mathbf{r}}{r}V_t({\mathbf r},{\mathbf 0})Q_{({\mathbf 0},t)}\right]^{\dagger}
\overline{Q}_{({\mathbf r},0)}
\frac{\boldsymbol{\gamma}\cdot\mathbf{r}}{r}V_0({\mathbf r},{\mathbf 0})Q_{({\mathbf 0},0)}\right\}
\right|0\right\rangle_U\\\nonumber
&=&2e^{-2m_Qt}\left\langle\tr\left\{V_t^{\dagger}({\mathbf r},{\mathbf 0})U_{{\mathbf r}}(t,0)
V_0({\mathbf r},{\mathbf 0})U^{\dagger}_{{\mathbf 0}}(t,0)\right\}
\right\rangle_U\\\label{eq:wil11}
&=& e^{-2m_Qt}\www=2e^{-2m_Qt}\left\langle W({\mathbf r},t)\right\rangle_U
,
\end{eqnarray}
\end{widetext}
where $\langle O\rangle_U$ denotes the expectation value of $O$ over gauge
configurations.

The trace above is over colour only. Hence the normalization of the Wilson
loop is, $\left\langle W({\mathbf r},0)\right\rangle_U=3$.  ${\mathcal
  T}=e^{-a{\mathcal H}}$ denotes the transfer operator and ${\tau}=t/a$ the
Euclidean time difference in lattice units.  The factor {\em two} originates
from $\tr\{ P_+\gamma_iP_-\gamma_j\}= \delta_{ij}\tr P_+^2=2\,\delta_{ij}$.

Note that in the first step of the derivation, a minus sign from the
commutator $\left[\overline{Q}_1\gamma_iQ_2\right]^{\dagger}
=-\overline{Q}_2\gamma_iQ_1$ is cancelled when permuting one quark field
through the remaining three others, prior to the Wick contraction.

Expectation values and correct pre-factors are understood to be implicit in
the pictorial representation of the correlators. In this case the time
direction is assigned to be vertical, the spatial separation to be
horizontal.  Light quark
propagators will be represented as wiggly lines, static quark propagators and
gauge transporters are shown as straight lines.

Next we consider the transition element between $\overline{Q}Q$
and $B_i\overline{B}_i$ states:
\begin{widetext}
\begin{eqnarray}
\nonumber
\left\langle B_i|{\mathcal T}^{\tau}|Q\right\rangle_U&=&
\left\langle 0\left|
T\left\{\overline{Q}_{({\mathbf 0},t)}\gamma_5{q}^i_{({\mathbf 0},t)}
\bar{q}^i_{({\mathbf r},t)}\gamma_5Q_{({\mathbf r},t)}
\overline{Q}_{({\mathbf r},0)}
\frac{\boldsymbol{\gamma}\cdot\mathbf{r}}{r}
V_0({\mathbf r},0)Q_{({\mathbf 0},0)}\right\}
\right|0\right\rangle_U\\\nonumber
&=&e^{-2m_Qt}\left\langle\Tr\left\{P_-\frac{\boldsymbol{\gamma}\cdot\mathbf r}{r}
M^{-1}_{({\mathbf 0},t);({\mathbf r},t)}U_{{\mathbf r}}(t,0)
V_0({\mathbf r},{\mathbf 0})U^{\dagger}_{{\mathbf 0}}(t,0)\right\}
\right\rangle_U\\\label{eq:wil12}
&=&e^{-2m_Qt}\wwb=e^{-2m_Qt}\wbw
=\left\langle Q|{\mathcal T}^{\tau}|B_i\right\rangle_U,
\end{eqnarray}
\end{widetext}
where the trace above is over colour and Dirac indices
and we have made use of the
relations, $\gamma_5P_+=P_-\gamma_5$, $\gamma_i\gamma_5=-\gamma_5\gamma_i$,
$\gamma_iP_+=P_-\gamma_i$ and $P_-^2=P_-$.

Finally, the $B_i\overline{B}_i$ sector reads:
\begin{widetext}
\begin{eqnarray}
\nonumber
\left\langle B_i|{\mathcal T}^{\tau}|B_j
\right\rangle_U&=&
\left\langle 0\left|
T\left\{\overline{Q}_{({\mathbf 0},t)}\gamma_5{q}^i_{({\mathbf 0},t)}
\bar{q}^i_{({\mathbf r},t)}\gamma_5Q_{({\mathbf r},t)}
\overline{Q}_{({\mathbf r},0)}
\gamma_5q^j_{({\mathbf r},0)}\bar{q}^j_{({\mathbf 0},0)}
\gamma_5Q_{({\mathbf 0},0)}\right\}
\right|0\right\rangle_U\\\nonumber
&=&e^{-2m_Qt}\delta_{ij}
\left\langle\Tr\left\{P_-M^{-1}_{({\mathbf r},t);
({\mathbf r},0)}U_{{\mathbf r}}(t,0)\right\}
\Tr\left\{P_+M^{-1}_{({\mathbf 0},0);
({\mathbf 0},t)}U^{\dagger}_{{\mathbf 0}}(t,0)\right\}\right\rangle_U\\\nonumber
&&-e^{-2m_Qt}\left\langle\Tr\left\{P_-M^{-1}_{({\mathbf 0},t);
({\mathbf r},t)}U_{{\mathbf r}}(t,0)P_+
M^{-1}_{({\mathbf r},0);({\mathbf 0},0)}U^{\dagger}_{{\mathbf 0}}(t,0)\right\}
\right\rangle_U\\\label{eq:wil22}
&=&e^{-2m_Qt}\left(\delta_{ij}\wbbd-\wbbc\right).\label{eq:wil2}
\end{eqnarray}
\end{widetext}
Again, the traces are over colour and Dirac indices.  In what follows, we will
also refer to the flavour singlet sector (which is the one relevant for
the string breaking problem) as the $I=0$ sector while we label
flavour non-singlet states as $I=1$. In the isospin $I=0$ sector
the connected diagram always contributes while the disconnected diagram
only contributes for $i= j$. In contrast, within the $I=1$ sector there is no
connected diagram but only the disconnected contribution\footnote{
If we allow the heavy quarks to move, by adding a kinetic term within a Born
Oppenheimer approximation, then the $I=0$ sector can be related to transitions
between vector bottomonium (or $0^{++}, 2^{++}, \ldots$ bottomonia)
into a pair
of $B$ and $\overline{B}$ pseudoscalar mesons or into $B^*$ and
$\overline{B}^*$ vector mesons (which are mutually degenerate in mass in the
static limit), with relative angular momentum $L$ chosen appropriately.  The
simplest example for this sort of process is $\Upsilon(4S)\rightarrow
B\overline{B}$ with final state $L=1$.  Within the $I=1$ sector one can write
down a similar mixing problem. In this case, the other state would be a
$\overline{Q}Q$ plus an $I=1$ meson, like the $\pi$.
The simplest such transition is
$B\overline{B}\rightarrow \eta_b+\pi$.}.

\subsection{Reduction to a $2\times 2$ matrix}
\label{sec:mix2}
We consider the scenario with $n_f>1$ mass-degenerate quark flavours.  By
summing over the flavour indices $i$ and $j$ in the above equations, the
correlation matrix can effectively be reduced to a $2\times 2$ problem: one can
easily orthogonalize the ${B}_i\overline{B}_i$ meson-meson states; for given
$r$ and $t$ all correlators, $\langle B_i|{\mathcal
  T}^{\tau}|B_j\rangle_U$, only involve one or two (for $i=j$)
generalized Wilson loops [which are displayed on the right hand side of
Eq.~(\ref{eq:wil2})]. For $n_f=2$ one can define for instance,
\begin{eqnarray}
\label{eq:mixi}
|B\rangle&=&\frac{1}{\sqrt{2}}\left(|B_1\rangle+|B_2\rangle\right),\\
|B_a\rangle&=&\frac{1}{\sqrt{2}}\left(|B_1\rangle-|B_2\rangle\right).
\label{eq:mixii}
\end{eqnarray}

Obviously $|B_a\rangle$ decouples from the other states:
\begin{eqnarray}
\langle Q|{\mathcal T}^{\tau}|B_a\rangle_U&=&\langle B|{\mathcal T}^{\tau}|
B_a\rangle_U=0,\\
\langle B_a|{\mathcal T}^{\tau}|B_a\rangle_U&=&e^{-2m_Qt} \wbbd.
\end{eqnarray}

This pattern easily generalizes to $n_f>2$: as soon as two or more indices are
antisymmetrized, the overlap with the $\overline{Q}Q$ state vanishes. Only the
completely symmetric
state has a non-trivial mixing (we write the formulas for general $n_f$
and $|B\rangle=\frac{1}{\sqrt{n_f}}\sum_{i=1}^{n_f}|B_i\rangle$):
\begin{eqnarray}
\langle Q|{\mathcal T}^{\tau}|B\rangle_U&=&\sqrt{n_f}e^{-2m_Qt}\wbw\\
\langle B|{\mathcal T}^{\tau}|B\rangle_U&=&e^{-2m_Qt}\left(\wbbd
-n_f\wbbc\right).
\end{eqnarray}

Hence, we have reduced the mixing problem for a general $n_f$
to a $2\times 2$
correlation matrix with elements,
\begin{equation}
C_{\alpha\beta}(t)
=\langle \alpha|{\mathcal T}^{t/a}| \beta\rangle_U,\quad
\alpha,\beta\in\{Q,B\}.
\end{equation}
This leads us
to the form already anticipated in Eq.~(\ref{eq:sb}),
\begin{equation}
\label{eq:fullcorr}
C(t)=e^{-2m_Qt}\left(\begin{array}{rl}
\quad \www&\quad \sqrt{n_f}\wwb\\&\\
\sqrt{n_f}\wbw&\quad -n_f\wbbc+\wbbd\end{array}\right),
\end{equation}
with the pictorial representations as defined
in Eqs.~(\ref{eq:wil11}) -- (\ref{eq:wil22}). 

Note that we have some freedom to change the normalizations, without affecting
the mass spectrum:
\begin{eqnarray}
C_{QB}\mapsto ab^*C_{QB},\quad 
C_{BQ}\mapsto a^*b\,C_{BQ},\\
C_{QQ}\mapsto |a|^2C_{QQ},\quad
C_{BB}\mapsto |b|^2C_{BB}.
\end{eqnarray}
The phase of the $C_{QB}=C_{BQ}^*$ element is irrelevant and we have employed
one of the two possible real choices. $C_{BB}=C_{BB}^{\rm   dis}-C_{BB}^{\rm
  con}$ consists of the following
disconnected and connected contributions, 
\begin{equation}
C_{BB}^{\rm dis}=e^{-2m_Qt}\wbbd\quad,\quad
C_{BB}^{\rm con}=n_fe^{-2m_Qt}\wbbc\quad.
\end{equation}

In what follows we will set $m_Q=0$, corresponding to a shift in all energy
eigenvalues. Differences between two energy levels, such as between the mass
of the $\overline{Q}Q$ system and twice the static-light mass, do not depend
on $m_Q$ and have a well defined continuum limit.  We remark, however, that
the levels themselves become cut-off independent only in the framework
of effective field theories. In this case, $m_Q(a)$ is required to cancel the
static self energy divergence and only the sum of quark masses and the
potential
is a ``physical'' quantity.

\section{Measurement Techniques}
\label{sec:latt}
We discuss the run parameters and the geometrical setup of our
simulation, before we elaborate on the noise reduction
and all-to-all propagator techniques
that we apply. We conclude by introducing the notations
that we will use in the interpretation of our numerical data.
\subsection{Simulation set-up}
\label{sec:simul}
We base our simulations on the $24^3\times 40$ T$\chi$L
configurations~\cite{Eicker:1998sy} with Wilson fermions
at $\kappa=0.1575$ and $\beta=5.6$. This translates
into $r_0=6.009(53)a$, corresponding to $a^{-1}\approx 2.37$~GeV or $a\approx
0.083$~fm from $r_0=0.5$~fm. The value of $r_0$  differs somewhat from the
earlier result, $r_0=5.892(27)a$~\cite{Bali:2000vr}, that we obtained without
accounting for mixing effects.  While our results on $C_{QQ}(t)$ are much more
precise than in this earlier study, after diagonalizing the mixing matrix,
the final
errors of the ground state energy level at $r< r_0$ increase.
At larger $r$, however, we achieve unprecedented
precision.  One obtains~\cite{Eicker:1998sy} $m_{\pi}a=0.276(5)$ and
$m_{\pi}/m_V=0.704(5)$, which means that the sea quark mass is slightly
smaller than that of the physical strange quark.

In order to stay clear of finite size effects, in particular within
the $B\overline{B}$ sector, it is advisable to
place the colour sources off-axis. An on-axis string breaking study,
in which ${\mathbf r}=n (1,0,0)a$, $n$ integer, would require a
spatial lattice extent $La>2\,r_c$, $r_c$ being the string breaking distance.
Off-axis separations allow for a relaxation of the above condition to
$La>2/\sqrt{3}\,r_c$ for the spatial diagonal, ${\mathbf r}=n(1,1,1)a$,
and $La>2/\sqrt{2}\,r_c$ along the planar diagonal, ${\mathbf r}=n(1,1,0)$.

We have performed measurements on the following set of geometries:
\begin{eqnarray*}
{\mathbf r}&=&n(1,0,0)a, n\leq 11,\\
{\mathbf r}&=&n(1,1,0)a, n\leq 11,\\
{\mathbf r}&=&n(1,1,1)a, n\leq 11,\\
{\mathbf r}&=&(10,10,n)a, n\in\{2,3,4,5,6\}\\ 
{\mathbf r}&=&(10,8,7)a,\\
{\mathbf r}&=&(10,9,7)a,
\end{eqnarray*}
as well as for ${\mathbf r}={\mathbf 0}$.
The distance resolution is enhanced around $r_c\approx 15\,a$,
to 10 points inside the range,
$14.14\approx 10\sqrt{2}\leq r/a\leq
15.59\approx 9\sqrt{3}$.
In order to increase statistics we average over
equivalent permutations and reflections of the axes.  We do not find any
directional dependence, even for the $(11,0,0)$ and $(11,11,0)$ points, and
hence there is no sign of finite size problems close to $r_c$.
In the neighbourhood of $r_c$, the largest component that
we employ is $r_i=10a$.

With $r_c\approx 15\,a$ and $L=24$
we obtain, $\sqrt{2}L/(2\,r_c)\approx
1.13>1$ and $\sqrt{3}L/(2\,r_c)\approx 1.39>1$. Therefore, the T$\chi$L
physical lattice extent $La\approx 2.0$~fm is sufficiently large
for our purpose. We remark that 
the pion correlation length also fits well into the
spatial lattice extent, $La>6\,m_{\pi}^{-1}$.

We extract the elements of $C(t)$ [Eq.~(\ref{eq:fullcorr})] that involve light
quark propagators from a set of 20 thermalized gauge configurations $\{
{\mathcal U}_i\} $, $i=1,\ldots,20$, separated by 125 Hybrid Monte Carlo
trajectories.  Earlier studies~\cite{Bali:2000vr,Lippert:1997qy,Bali:2001gk}
have established that these configurations are effectively independent of each
other.

The standard Wilson loop $C_{QQ}(t)$ is determined on a larger ensemble of
184 configurations, separated by 25 trajectories.  We also wish to eliminate
possible autocorrelations in this case.  Moreover, we attempt to consistently
take account of correlations between different matrix elements (that have been
determined on one and the same set of configurations).  To this end, the 184
configurations are averaged into 20 bins that are mapped onto the ensemble
$\{{\mathcal U}_i\}$. Each bin $i$ contains the five configurations that are
closest in Monte Carlo time to the above mentioned
20 configurations as well as an
additional four to five configurations from within another region of the time
series. As it turned out, the limiting factor of our statistical resolution
is the accuracy of the Wilson loop data and hence little can be gained from
increasing our sample size for $C_{QB}(t)$ and $C_{BB}(t)$ beyond 20
configurations.

\subsection{Signal enhancement techniques}
\label{sec:over}
We are interested in the exponential decay of the elements of the correlation
matrix $C(t)$ at large Euclidean times.
Statistically significant results cannot be achieved unless
the asymptotic behaviour can already be extracted at moderate time separations.
To this end, we employ smearing techniques that enhance the overlap of
the operators used in the creation of particular states with the corresponding
physical ground states, without affecting the eigenvalues.
Furthermore, the noise/signal ratio has to be controlled. In pure gauge
theories extended operators can be constructed, with reduced variance,
retaining
identical expectation values~\cite{Luscher:2001up,Parisi:1983hm}.
Unfortunately, these techniques, which exploit the locality of the gauge action
in space-time,
are not applicable when including sea quarks that (after integration)
induce non-localities. Instead of reducing the variance,
we enhance the signal by
an appropriate choice of the lattice static quark action.

\subsubsection{Smearing}
\label{sec:smear}
We employ the iterative APE~\cite{Albanese:1987ds,Teper:1987wt}
smearing procedure for the spatial transporters $V_t$ that enter
the creation operator
of the $\overline{Q}Q$ states, Eq.~(\ref{eq:gamma}):
\begin{equation}
\label{eq:smear}
U_{x,i}^{(n+1)}= P_{SU(3)}\left(U_{x,i}^{(n)}+\alpha\sum_{|j|\neq i}
U_{x,j}^{(n)}U^{(n)}_{x+a\hat{\boldsymbol{\jmath}},i}U^{(n)\dagger}_{x+a\hat{\boldsymbol{\imath}},j}\right),
\end{equation}
where $i\in\{1,2,3\}, j\in\{\pm 1,\pm 2,\pm 3\}$.
$P_{SU(3)}$ denotes a projection operator, back onto the $SU(3)$ group
and the sum is over the four spatial ``staples'', surrounding $U_{x,i}^{(n)}$.
After extensive studies we employ the parameter values
$N_{\mbox{\scriptsize APE}}=50$ for the number of smearing iterations and
the weight factor $\alpha=2.0$. For the projection operator we
somewhat deviate from
Ref.~\cite{Bali:2000vr} where we maximized
$\mbox{Re}\,\tr\{ A^{\dagger} P_{SU(3)}(A)\}$,
iterating over $SU(2)$ subgroups. Instead we define,

\begin{eqnarray}
A'=\frac{A}{\sqrt{A^{\dagger}A}}\in U(3),\\
P_{SU(3)}(A)=A'\det(A')^{-1/3}.\label{eq:phase}
\end{eqnarray}
The inverse square root is calculated in the (orthogonal) eigenbasis
of $A^{\dagger}A$, where we take the positive root of the respective
(positive) eigenvalues. In general there are three possible choices
for the phase correction, Eq.~(\ref{eq:phase}). We take the one that is closest
to unity. Note that this construction guarantees (except in singular
cases that we never encountered in numerical simulations)
$[P_{SU(3)}(A)]^{-1}=[P_{SU(3)}(A)]^{\dagger}$ as well as gauge covariance:
$UP_{SU(3)}(A)V=P_{SU(3)}(UAV)$ for $U,V\in SU(3)$.

Subsequently, we construct the spatial transporters $V_t$ by calculating
products of the APE smeared links along paths that stick to the direct
connection between quark and antiquark as closely as possible. In this way,
the overlap between creation operator and physical $\overline{Q}Q$ state
is vastly enhanced.

We improve the overlap of our mesonic operators with the
static-light ground state by applying
Wuppertal smearing~\cite{Gusken:1989ad},
\begin{equation}
\label{eq:wuppertal}
\phi^{(n+1)}_x=\frac{1}{1+6\delta}\left(\phi^{(n)}_x+
\delta\sum_{j=\pm 1}^{\pm 3}U_{x,j}\phi^{(n)}_{x+a\hat{\boldsymbol{\jmath}}}\right),
\end{equation}
to light quark fields $\phi$, where we set $\delta=4$ and replace
$U_{x,j}$ by the APE smeared links as detailed above.
We then employ
the linear combination $\phi^{(20)}-6.6323\phi^{(40)}+7.2604\phi^{(50)}$ as
our smearing function.
Note that we are calculating local-local all-to-all propagators
to which we can subsequently apply Wuppertal smearing.

Best results are
obtained by using smeared-local quark propagators.
For a single static-light meson positivity of the coefficients
in the spectral decomposition is not guaranteed.
Neither do we recover positivity for the bound $B\overline{B}$ system
as the source is smeared at position ${\mathbf 0}$ while the sink
is smeared at position ${\mathbf r}$: the wave function is not
symmetrized with respect to ${\mathbf 0}\leftrightarrow {\mathbf r}$.

\subsubsection{Static quark action}
\label{sec:statq}
One problem in simulations with static sources is the rapid exponential
decay of the associated Green functions with Euclidean time. One of the
reasons for this is a large static quark self energy contribution
which, to leading order in perturbation theory, reads,
$\delta m\approx c\,C_F\,\alpha_L\,a^{-1}$, with a constant $c\approx
1.587956$.
This contribution obviously diverges with $a^{-1}$.

There is however some freedom in the choice of
the static action, i.e.\ in the choice of a lattice discretization
of $D_4$ within Eq.~(\ref{eq:static}), as long as the action
remains localized and converges towards the continuum action in
the limit $a\rightarrow 0$. This choice will affect
the lattice definition of the Schwinger line Eq.~(\ref{eq:schwinger}).
One possible such discretization reads,
\begin{equation}
D_4Q_x = a^{-1}\left(Q_x-\overline{U}^{\dagger}_{x-a\hat{4},4}Q_{x-a\hat{4}}\right),
\end{equation}
with
\begin{equation}
\label{eq:fat}
\overline{U}_{x,4}=P_{SU(3)}\left(\epsilon\, U_{x,4}+\sum_{j=\pm 1}^{\pm 3}
U_{x,j}U_{x+a\hat{\boldsymbol{\jmath}},4}U^{\dagger}_{x+a\hat{4},j}
\right),
\end{equation}
where we use $\epsilon=0$.
Note that this procedure is reminiscent of APE smearing,
Eq.~(\ref{eq:smear}), 
but with a sum over six rather than over four
staples. 

The Schwinger line that appears within the corresponding
static propagator can now be written as,
\begin{equation}
U_{{\mathbf x}}(t,0)=T\,\prod_{\tau=0}^{t/a-1}
\overline{U}_{({\mathbf x},\tau a),4}.
\end{equation}
$T$ denotes the time ordering operator.

The ``extended'' temporal links correspond to introducing
``form factors'' in perturbation theory~\cite{Capitani:2002mp}:
to leading order, replacing $U_{x,4}$ by $\overline{U}_{x,4}$
with the (optimal)
weight $\epsilon=0$ is equivalent to multiplying the self energy by a factor
$\approx 1/2.94$: $c\mapsto c-\pi/3$. The signal is exponentially improved in
$t$, while  the  absolute noise approximately
maintains its level. Since the self energy cancels
from energy differences as well as from the sum of $2m_Q$ plus
energy levels, the physics of string breaking remains unaffected.
Only at small distances, we encounter different lattice terms, which
(being artefacts of the discretization) do not alter the continuum
limit.
Fat temporal links can also influence 
the ground state overlaps:   our   impression is that
they help to improve the situation further.

One can define a tree level improved lattice
distance~\cite{Sommer:1993ce,Necco:2001xg},
\begin{equation}
\label{eq:rbar}
\overline{r}=r\left[1+O(a^2)\right]=a\left[\frac{1}{({\mathbf r}/a)}\right]_L^{-1},
\end{equation}
where $[1/{\mathbf R}]_L\rightarrow 1/R$ for $R\rightarrow\infty$
denotes the tree level [$O(\alpha_s)$]
lattice propagator in an appropriate normalization.
It can easily be shown that replacing thin temporal
links by fat temporal links, Eq.~(\ref{eq:fat}), only affects
$[1/{\mathbf R}]_L$ at distances ${\mathbf R}\in
\{{\mathbf 0},\hat{\boldsymbol{\imath}}\}$. Amusingly, iterating the
``fattening'' $n$ times will leave the tree-level expressions
at all distances $R_1+R_2+R_3>n$ invariant.
Note that $\overline{0}=1/(2c)\approx 0.31a (0.92a)\neq 0$
and $\overline{a}\approx 0.92a (1.37a)\neq a$
for standard (fat) temporal links.
In principle, one could further fatten temporal links but in view of
increasing the small-$r$ distortions and of the reduction of the static energy already
achieved, to one third of its original value, we refrain from doing so.
Throughout this paper we will plot all $r$-dependent data as
functions of $\overline{r}$, thus removing the short-distance
lattice direction-dependence to $O(\alpha_s)$. 

In the present context, fat temporal links have first been
employed in Ref.~\cite{Hasenfratz:2001hp}. In this case,
more refined actions were implemented, utilising all non-intersecting
paths that can be constructed within an elementary hypercube.
Other studies employing similar techniques can be found in
Refs.~\cite{Bernard:2002sb,Bornyakov:2004ii,DellaMorte:2003mn,Okiharu:2004tg}.

\begin{figure}[th]
\hspace{-0.1in}
\epsfxsize=0.9\columnwidth
\centerline{
\epsffile{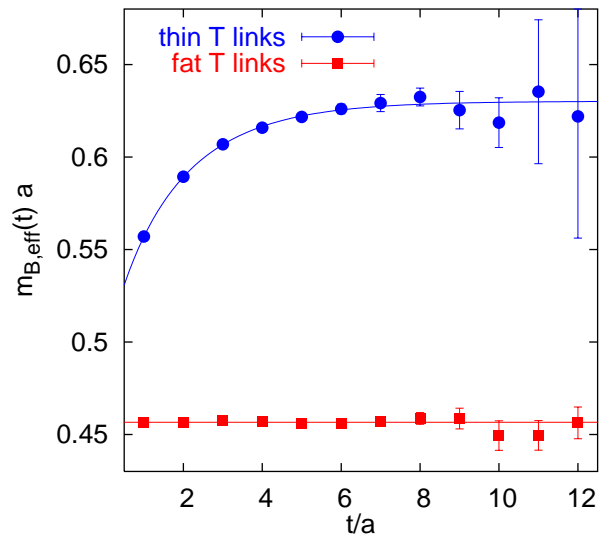}}
\caption {Comparison of effective masses of static-light correlation
functions, obtained employing static actions with and without fat temporal
links. The wave function has been optimized to yield best ground state overlap
for the fat link static action. SET and HPA have been applied in both cases.}
\label{fig:tsmear1}
\end{figure}
In Figure~\ref{fig:tsmear1} we plot effective masses,
\begin{equation}
\label{eq:effmass}
m_{X,\mbox{{\scriptsize eff}}}(t)=a^{-1}\ln\frac{C_X(t)}{C_X(t+a)},
\end{equation}
of static-light mesons
with and without fat time links. $C_B(t)$ stands
for the static-light correlation function. The curves
correspond to one- and two-exponential fits to the fat and thin link
data, respectively. 
Note  that the use of SET and HPA techniques (described in
Sec.~\ref{sec:alltoall} below)  was essential for achieving 
the high signal quality.
The new static action
shifts the mass by an amount, $\delta m= 0.174(7)\,a^{-1}$.
The absolute statistical errors of the correlation function increase somewhat,
however, the relative statistical errors are reduced, in particular at
large times as the signal falls off less steeply. This in turn results
in a reduction of the error of the effective masses, in particular at
large $t$. The Figure also illustrates that we are able to achieve
an excellent overlap with the physical ground state.

\begin{figure}[th]
\hspace{-0.1in}
\epsfxsize=0.9\columnwidth
\centerline{
\epsffile{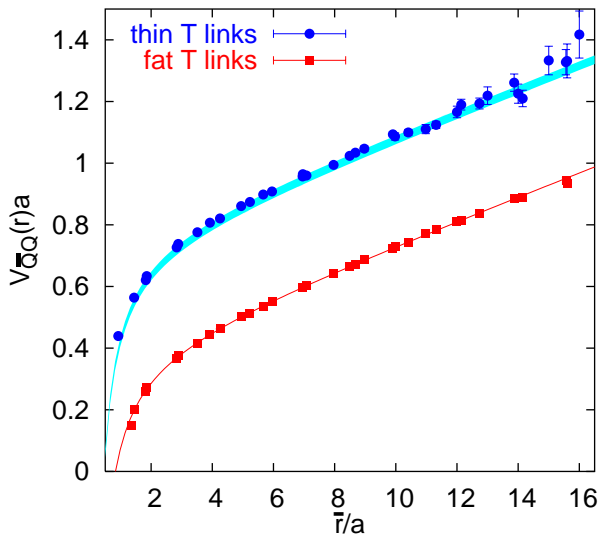}}
\caption {Comparison of the $\overline{Q}Q$ potential in lattice schemes
with and without fat temporal links, in both cases for
$t_{\min}/a=5$. String breaking is expected to take place around
$\overline{r}=r_c\approx 15\,a$ but this is not visible from the Wilson loop
data alone. The curve represents a funnel fit to the fat link data
and the error band is this parametrization shifted upwards by
the amount $2\delta m$, where $a\delta m=0.174(7)$.}
\label{fig:tsmear2}
\end{figure}

In Figure~\ref{fig:tsmear2} we compare the static potentials as calculated
from the Wilson loop operator $C_{QQ}(t)$ alone,
with and without fat time links. These potentials have been obtained
from single exponential fits with $t_{\min}/a=5$.
The curve corresponds to a funnel-type $r$-dependence with fit range
$4a\leq\overline{r}\leq 13a$. String breaking is expected around
$\overline{r}=r_c\approx 15\,a$. However, this is not seen in
the data. The error band represents our expectation
for the thin link potential, obtained from the respective static-light
mass shift $\delta m$, as determined above. We find consistency.

The extended static action leaves the
absolute errors of the correlation functions
basically constant but still improves the signal exponentially.
As a result, the effective mass errors are reduced impressively,
by a factor of about {\em five} throughout.\footnote{
At first sight,
the comparatively modest improvement of the static-light data  seems to be
in contradiction to the very 
significant effects observed in Ref.~\cite{DellaMorte:2003mn}. However,
a closer inspection reveals that without employing
our additional improvement methods,
i.e.\ averaging over all lattice points by means of stochastic estimates
and employing HPA,
the gain factor from using the extended static action would have been larger:
the signal over noise improvement
appears to saturate, after adding more and more tricks.}.

\subsection{All-to-all propagator techniques}
\label{sec:alltoall}
The requirement of quark propagators from all
source to all sink locations is obvious:
within  the $C_{BB}$ element of the correlation matrix,
Eq.~(\ref{eq:fullcorr}), we encounter light quark propagators
starting from different source
positions.
Moreover, we can reduce the notorious noise levels of disconnected
(and of some connected) diagrams;
all-to-all propagators allow for the full exploitation
of translational invariance, increasing the accuracy of the entire
correlation matrix.

Since the propagator $M^{-1}$, Eq.~(\ref{eq:wilferm}), has
$12 V\times 12V$ components (in our case $V=24^3\times 40$), direct
evaluation would be prohibitively expensive, both in terms of
computer time and of memory. However, the correct result can
also be obtained by combining the truncated eigenmode
approach~\cite{Neff:2001zr,Schilling:2004kg,Setoodeh:1988ds}
(TEA) with stochastic estimator techniques (SET).
Where possible, we improve both,
the convergence of TEA and the statistical errors of
SET, by employing the hopping
parameter acceleration (HPA).

For completeness we introduce   these
three techniques (TEA, SET and HPA)
and their implementation  in the following subsections. We
conclude by
comparing numerical data obtained by use of combinations of these methods.

\subsubsection{Truncated eigenmode approach (TEA)}
The fermionic propagator is the inverse of the Wilson Dirac matrix
$M$ of Eq.~(\ref{eq:wilferm}). However, $M$ is not Hermitian which is why
we define,
\begin{equation}
Q=\gamma_5M.
\end{equation}
The relation $M^{\dagger}=\gamma_5M\gamma_5$ implies
Hermiticity of $Q=Q^{\dagger}$.
We calculate the
smallest $n=200$ (real) eigenvalues $q_i$ and corresponding
orthonormal eigenvectors $|u_i\rangle$, $i=1,\ldots, n$,
\begin{equation}
Q|u_i\rangle=q_i|u_i\rangle\quad,\quad
\langle u_i|u_j\rangle=\delta_{ij}.
\end{equation}

This is done by means of the parallel implicitly restarted
Arnoldi method (IRAM) with Chebychev acceleration~\cite{Neff:2001zr},
using the PARPACK library~\cite{arnoldi}.
We can now approximate,
\begin{equation}
\label{eq:tea}
Q^{-1}\approx\sum_{i=1}^n|u_i\rangle q_i^{-1}\langle u_i|.
\end{equation}
Obviously, $M^{-1}=Q^{-1}\gamma_5$.
With this truncated eigenmode approach (TEA), we have reduced a $12 V\times 12 V$ problem
to a $12 V\times 12 n$ problem. 

Another nice feature of this procedure is that there is no critical
slowing down of the algorithm as the quark mass is reduced.
However, the difference between the left and
right hand sides of Eq.~(\ref{eq:tea}) is systematic. One can in
principle estimate this bias from the convergence properties under
variation of $n$~\cite{Neff:2001zr}.
In the present context we will
render the result {\em exact} by stochastically estimating
the remainder, replacing the systematic error by a statistical uncertainty.
For this purpose, it is useful to define the projection
operator ${\mathcal P}_n$, onto the basis
spanned by the first $n$ eigenvectors,
\begin{equation}
{\mathcal P}_n=\sum_{i=1}^n|u_i\rangle\langle u_i|.
\end{equation}
Note that for $n$ smaller than the rank of $Q$,
this basis is truncated and hence incomplete:
${\mathcal P}_n\neq{\mathbf 1}$. However, $[Q,{\mathcal P}_n]=0$.
We can also define the projector onto the orthogonal
subspace, ${\mathbf 1}-{\mathcal P}_n$.

\subsubsection{Stochastic estimator techniques (SET)}
Stochastic estimator  techniques have been applied by various groups in the
past~\cite{Bitar:bb,Bernardson:yg,Dong:1995rx,Eicker:1996gk,Thron:1997iy,Wilcox:1999ab,Michael:1999nq,Struckmann:2000bt,McNeile:2004wu}.
We introduce the following notation,
\begin{equation}
\overline{O}=\frac{1}{N}\sum_{j=1}^NO^j.
\end{equation}
$N$ denotes the number of ``stochastic
estimates''.
Let $|\eta^i\rangle$, $i=1,\ldots,N$ be random vectors with the
properties,
\begin{eqnarray}
\overline{|\eta\rangle}= O(1/\sqrt{N}),\\
\overline{|\eta\rangle\langle\eta|}= {\mathbf
1}+O(1/\sqrt{N}).\label{eq:cancel}
\end{eqnarray}
These requirements are for instance met
if the $12V$ components are numbers $e^{i\phi}$, with the uncorrelated
phases $\phi\in\{\pm\pi/4,\pm 3\pi/4\}$ randomly selected.
We employ such a complex ${\mathbb Z}_2$ noise,
where our
random vectors take values over the entire four-volume, flavour and colour.

If we solve the linear system,
\begin{equation}
\label{eq:lin}
Q|s^i\rangle=|\eta^i\rangle,
\end{equation}
for $|s^i\rangle$ then, for large $N$,
we can substitute [Eq.~(\ref{eq:cancel})],
\begin{equation}
\label{eq:set}
Q^{-1}\approx\overline{|s\rangle\langle\eta|}.
\end{equation}
Note that in our study we actually invert $A=M^{\dagger}M=Q^2$ and then obtain
$Q^{-1}$ by multiplying the solution with $Q$.
This allows for more flexibility: for instance the Roma
smearing technique~\cite{deDivitiis:1995pv},
which amounts to the replacement
$Q^{-1}\mapsto A^{-1}$ within hadronic
Green functions, can readily be implemented.
It can then be shown by means of spectral
decompositions that in many cases the ground state mass remains
unaffected~\cite{deDivitiis:1995pv}.
In the present context we have made use of this method, in addition to
standard smeared-smeared and smeared-local correlation functions,
within the optimization procedure of the
static-light creation operator. Unfortunately,
Roma-smearing turns out not to be applicable to the 
$B\overline{B}\leftrightarrow\overline{Q}Q$
mixing problem.

The sparse linear system of Eq.~(\ref{eq:lin}) is solved by means of the
BiCGstab2 algorithm~\cite{Frommer:vn}. 
Unlike in Eq.~(\ref{eq:tea}) where the bias was systematic,
the difference between the approximation of Eq.~(\ref{eq:set}) and
the exact result is purely statistical and reduces like $1/\sqrt{N}$.
In order to limit the computational effort,
$N$ should not be chosen overly large.
However, the noise level from SET should
at least match the one from the (finite) sampling of gauge configurations.
In general,
the optimal balance in both samplings will also depend on the observable
in question and on the methods employed.

We can estimate the difference between the TEA approximation and
the true result by means of SET.
The smaller this difference, the smaller the statistical errors
will be that are introduced by SET.
Hence TEA can be employed to reduce the variance of SET.
We project the right hand side
of Eq.~(\ref{eq:set}) into the subspace which is orthogonal to the
TEA eigenvectors:
\begin{equation}
(1-{\mathcal P}_n)Q^{-1}(1-{\mathcal P}_n)
\approx\frac{1}{N}\sum_{j=1}^N(1-{\mathcal P}_n)|s^j\rangle\langle
\eta^j|(1-{\mathcal P}_n).
\end{equation}
In practice this is done by calculating and storing,
\begin{eqnarray}
|\tilde{s}^j\rangle&=&|s^j\rangle-\sum_{i=1}^n|u_i\rangle\langle u_i|
s^j\rangle,\\
|\tilde{\eta}^j\rangle&=&|\eta^j\rangle-\sum_{i=1}^n|u_i\rangle\langle u_i|
\eta^j\rangle.
\end{eqnarray}
Then,
\begin{equation}
\label{eq:invQ}
M^{-1}\approx\left[\sum_{i=1}^n|u_i\rangle q_i^{-1}\langle u_i|+
\frac{1}{N}\sum_{j=1}^N|\tilde{s}^j\rangle\langle\tilde{\eta}^j|\right]\gamma_5.
\end{equation}
Note that there is no systematic error
on finite-$N$ approximants
but only a statistical $O(1/\sqrt{N})$ uncertainty. In the present context we
found $N=50$, combined with $n=200$, to suffice for calculating $C_{QB}(t)$
and static-light
meson correlators. Within the two diagrams contributing to
$C_{BB}(t)$, it is necessary to choose
two independent random sources as in either case there exist
the same two possibilities of connecting sources with sinks. In these cases we
calculate the SET corrections for the two respective propagators
independently, with $N=25$ random sources each. Subsequently,
we interchange the two sets of random sources to increase the
statistics at little computational overhead. Such ``recycling'' has 
been pioneered by the Dublin group~\cite{Cais:2004ww}.

As long as we are only interested in using SET to remove
the bias from TEA for a fixed
$n$ we can in principle solve Eq.~(\ref{eq:lin}) within the orthogonal
subspace only, substituting the random sources $|\eta^{j}\rangle$ on the
right hand side with
$|\tilde{\eta}^j\rangle$. We attempted this but found no advantage in terms
of real cost in computer time. This of course might change at smaller
quark masses or with different light quark actions.
However, having random source solutions at our disposal that are independent
of the TEA allows for more flexibility. For instance, not all physical
states will be dominated by the lowest lying eigenmodes of $Q=\gamma_5M$.
In particular, the TEA contribution to $C_{BB}^{\rm con}(t)$
turned out to be tiny, such that in the end we reduced the
cost to compute $C_{BB}^{\rm con}(t)$,
by employing a stand-alone SET.

Needless to say that, once we have calculated all-to-all propagators,
Wuppertal smearing, Eq.~(\ref{eq:wuppertal}), can
be implemented. In principle, one could even variationally
optimize the smearing function~\cite{Draper:1993qj}, for instance after
fixing to Coulomb gauge. However, our smearing function
turned out to be already so highly optimized
that further gain was too hard to achieve.
\subsubsection{Hopping parameter acceleration (HPA) of TEA and SET}
The main motivation of complementing SET with TEA is to reduce the
signal that needs  estimation and hence the stochastic
errors. One might ask if it is possible to further facilitate the low
eigenvalue dominance, accelerating the convergence of TEA (and of SET).
This is indeed possible by applying what we call the hopping parameter
acceleration (HPA).

We rewrite the fermionic matrix Eq.~(\ref{eq:wilferm}) as,
\begin{equation}
M={\mathbf 1}-\kappa D.
\end{equation}
For sufficiently small hopping parameter values $\kappa<\kappa_c$,
one can expand,
\begin{equation}
M^{-1}=\sum_{i=0}^{\infty}(\kappa D)^i=
\sum_{i=0}^{k-1}(\kappa D)^i+(\kappa D)^kM^{-1},
\end{equation}
where $k\geq 1$. The idea now is that for distances between source and
sink that are bigger than $k$ lattice spacings
the first term on the right hand side does not
contribute. This can
readily be 
seen as follows:
as $D$ only connects nearest space-time neighbours [and ${\bf 1}=(\kappa D)^0$
only contains diagonal entries],
the sum vanishes within
elements $M_{xy}^{-1}$ if $\sum_{\mu}|x_{\mu}-y_{\mu}|/a\geq k\geq 1$.
$M_{xy}^{-1}$
can be replaced by $[(\kappa D)^kM^{-1}]_{xy}$.
This means that $Q^{-1}_{xy}=[(\kappa D)^kQ^{-1}]_{xy}$.
With $M|r_i\rangle=\mu_i|r_i\rangle$,
$\langle l_i|M=\mu_i\langle l_i|$, where $\mu_i$ are the eigenvalues of
$M$,
Eq.~(\ref{eq:tea}) can be substituted by,
\begin{equation}
\label{eq:tea3}
Q^{-1}_{xy}=M^{-1}_{xy}\gamma_5\approx\sum_{i=1}^n\langle x|r_i\rangle (1-\mu_i)^k\mu_i^{-1}\langle l_i|y\rangle\gamma_5,
\end{equation}
where again $k$ is smaller or equal to the number of links separating source
from sink. 
For large $k$, contributions from big eigenvalues of $M$ are suppressed and
the expression becomes all the  more dominated by low
lying eigenmodes. Hopefully, the dominance in terms of low eigenmodes of
$M$ will then
also apply to low eigenmodes of $Q$.

In fact we do not only find HPA to improve TEA but the main effect of
HPA is with respect to SET\footnote{
Note that variance reduction methods, that make use of the hopping parameter
expansion, have been pioneered by the Kentucky group~\cite{Thron:1997iy},
in a different setting (see also~\cite{Wilcox:1999ab}).}: the cancellation
of stochastic noise is accelerated if the number of contributions to
the stochastic average is reduced. Short-distance noise is accompanied
by larger amplitudes than large-distance noise and hence its cancellation
requires a comparatively larger number of noise vectors. HPA explicitly
eliminates such short-distance contributions.

The benefit from HPA will increase with larger
temporal or spatial distances.
Unfortunately, in the limit of light quark masses, as $\kappa$ approaches
$\kappa_c$, the quark propagator will decay less
rapidly with the distance and the explicit treatment of the first
few terms within the hopping parameter expansion will have less
of an effect. This is also obvious from the reduced convergence
of the hopping
parameter expansion at small quark masses.
In this case we would however
expect a better convergence of the TEA contribution in the first place.

Note that HPA exploits the ultra-locality
of the Wilson action and does not generalize for instance to
the Neuberger action~\cite{Neuberger:1997fp}.
Again, this might be compensated for, by a faster convergent
TEA approximation, due to the improved chiral properties
of the chiral actions, in particular at small quark masses.
In contrast, the ``dilution'' method advocated in Ref.~\cite{Cais:2004ww}
will still be applicable in a setting with chiral fermions,
reducing the variance of SET
for the very same reasons as HPA does.

By applying HPA to the whole matrix, cf.\ Eq.~(\ref{eq:invQ}),
we exploit both effects, the improvement of the low eigenvalue dominance
and the variance reduction of SET:
\begin{equation}
M^{-1}\approx\left(\kappa D\right)^k\left[\sum_{i=1}^n|u_i\rangle q_i^{-1}\langle u_i|+
\frac{1}{N}\sum_{j=1}^N|\tilde{s}^j\rangle\langle\tilde{\eta}^j|\right]\gamma_5,
\end{equation}
where, as above, $ka$ is the lattice-distance between source and sink.
Again, the above equation is exact up to statistical $O(1/\sqrt{N})$
corrections.

Unfortunately, due to the size of our smearing function, we cannot
employ HPA for propagators along spatial separations, i.e.\ within
$C_{QB}(t)$ or within $C_{BB}^{\rm con}(t)$. However,
$C_{BB}^{\rm dis}(t)$ benefits from this technique
as do static-light correlation functions. One way of extending it
to the before-mentioned elements is to cut off the radius
of the smearing function. Such a cut-off, in conjunction with
an iterative smearing method, is hard to implement. One way out
would be to work in Coulomb gauge with fixed
weight-factors~\cite{Draper:1993qj}.
Another possibility is the implementation of a rotationally non-invariant
smearing function. But in this case
it turned out to be difficult to sustain an
acceptable ground state overlap.

\subsubsection{Comparative study of SET, TEA and HPA}
We demonstrate the impact of the above methods 
for the example of the static-light meson mass, in the scenario of the
fat link static action described in Sec.~\ref{sec:statq}.
We also verify the potential of  HPA for the example of $C_{QB}(t)$,
however, without smearing (see above).
\begin{figure}[th]
\hspace{-0.1in}
\epsfxsize=0.9\columnwidth
\centerline{
\epsffile{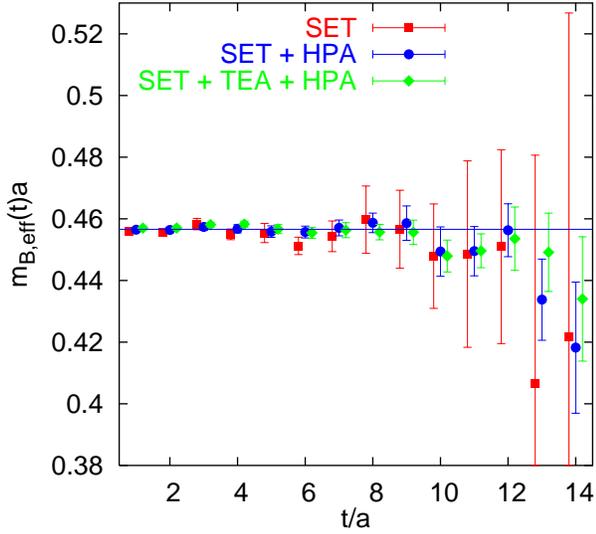}}
\caption {Effective static-light masses, obtained with SET alone,
with HPA SET and with all three methods combined.}
\label{fig:compare}
\end{figure}
\begin{figure}[th]
\hspace{-0.1in}
\epsfxsize=0.9\columnwidth
\centerline{
\epsffile{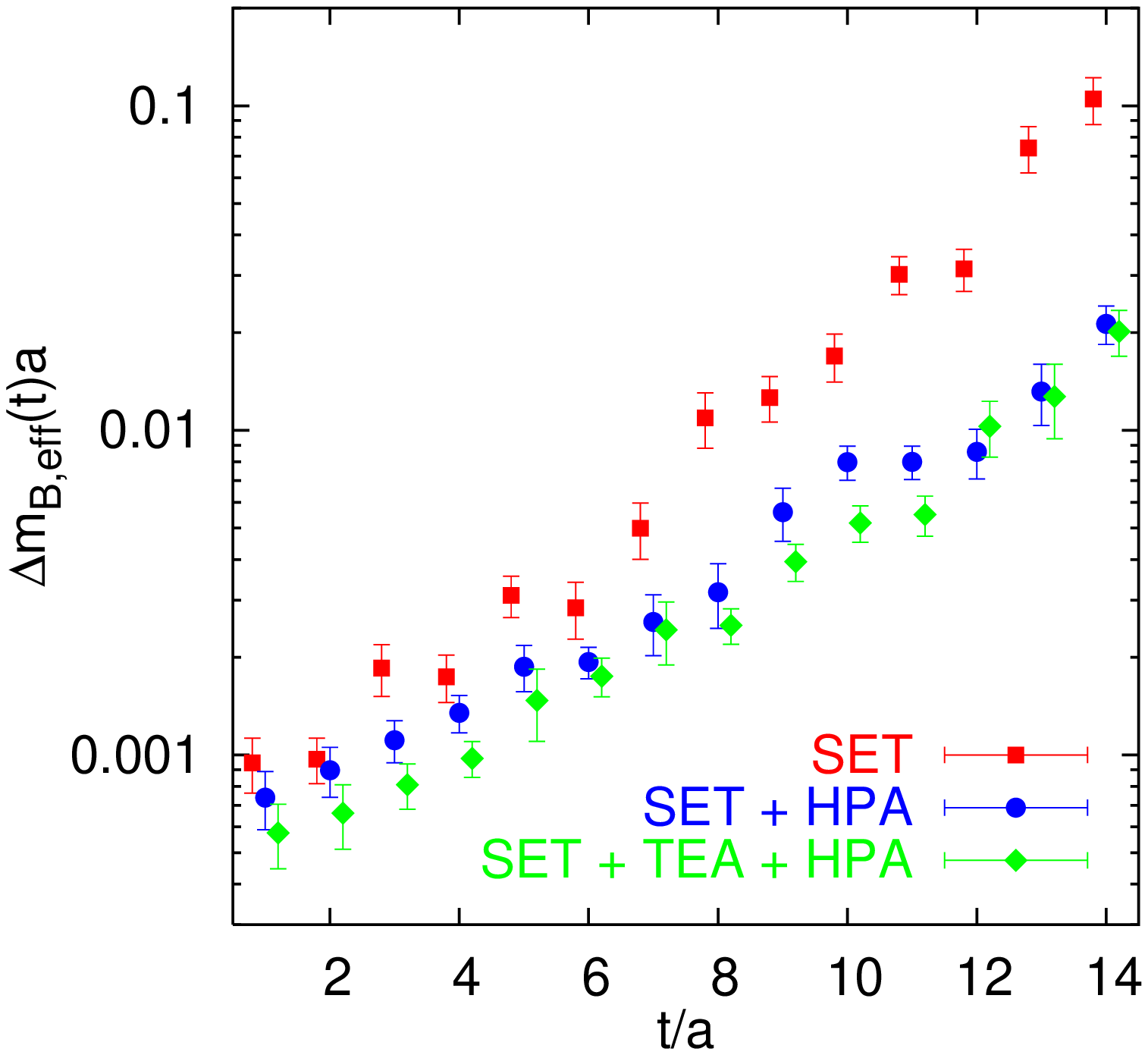}}
\caption {The errors of effective static-light masses, obtained with SET alone,
with HPA SET and with all three methods combined.}
\label{fig:compare2}
\end{figure}
\begin{figure}[th]
\hspace{-0.1in}
\epsfxsize=0.9\columnwidth
\centerline{
\epsffile{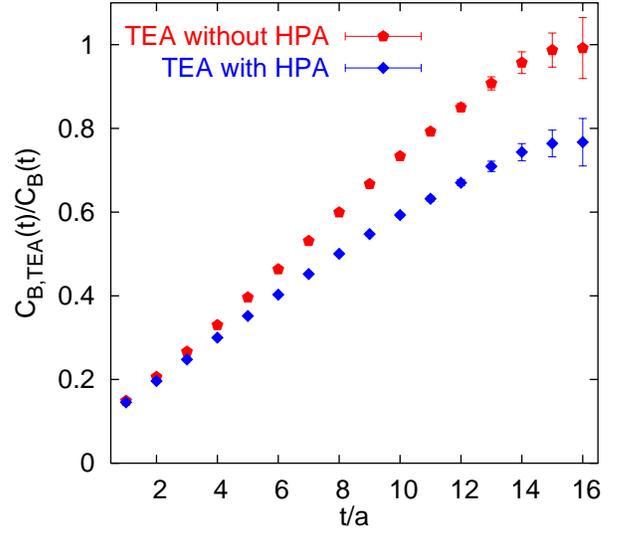}}
\caption {The relative magnitude of the TEA contribution within the
static-light correlation function, with and without HPA.}
\label{fig:compare3}
\end{figure}

In Figure~\ref{fig:compare} we display effective masses Eq.~(\ref{eq:effmass}),
obtained with SET alone as well as with HPA SET,
and after explicitly calculating the contribution
from the first 200 eigenmodes (TEA). Note that the ordinate covers
a huge $t$-range, up to a distance $t\approx 1.2$~fm.
Also note the magnified scale of the abscissa, covering
the window $0.9\,\mbox{GeV}<m_{B,{\rm eff}}<1.26$~GeV.
In particular at large
times, HPA impressively reduces the errors and
TEA results in some additional improvement. This is quantified in
Figure~\ref{fig:compare2}, where we display the respective statistical
errors themselves. Note the logarithmic scale. For instance, at $t=8a$
HPA reduces the SET error to about one third of its original
value while TEA yields another $\approx 20$~\% reduction. Since
the effective mass
is approximately independent of $t$ the absolute errors displayed are
proportional to the relative errors which (as is obvious from the
Figure) grow exponentially with $t$. Fortunately, both, HPA and TEA
reduce the amplitude and the exponent governing this
error increase, resulting in an exponential improvement at large times. 

In Figure~\ref{fig:compare3} we see
that the static-light correlation function at large times becomes
dominated by TEA.
Without HPA this dominance seems to be achieved earlier than with HPA:
the sea quark mass still appears too heavy for HPA to significantly
enhance
the low eigenvalue dominance of $Q=\gamma_5M$, for the correlation function
in question. However, the seemingly
perfect agreement for the stand-alone TEA case
of the displayed ratio at large $t$ with {\em one} is largely accidental.
Increasing the number of eigenvectors on one configuration
revealed that TEA tends to overshoot the exact result, before
converging towards it. HPA reduces this tendency.
Note that HPA cannot be applied to the SET part alone,
within the SET plus TEA combination.

In the HPA case TEA also results in an impressive reduction
of the signal that
remains to be estimated. However, a comparison between
SET plus HPA and SET plus HPA
plus TEA reveals that
the additional error reduction due to TEA
is only moderate.
After HPA the SET error is already at the level of the statistical fluctuations
between gauge configurations and of a comparable size to the
(non-stochastical) TEA error. In this situation,
substituting part of one signal by the other leaves the resulting
statistical error largely unaffected. This would have been different
for a larger statistical sample or at smaller sea quark masses.

Finally, we wish to investigate the effect of HPA on correlators
as a function of spatial source-sink separations. This is done for the example
of $C_{QB}(t)$, without smearing\footnote{
Our smearing function includes one
contribution with 50 Wuppertal smearing iterations which would mean that
the exponent $k$ has to be smaller than the source-sink link distance minus
49. Even for separations along a spatial diagonal,
HPA would only be applicable for distances much larger than half the lattice
extent. When using local sources and sinks we do not encounter such
restrictions.}.

\begin{figure}[th]
\hspace{-0.1in}
\epsfxsize=0.9\columnwidth
\centerline{
\epsffile{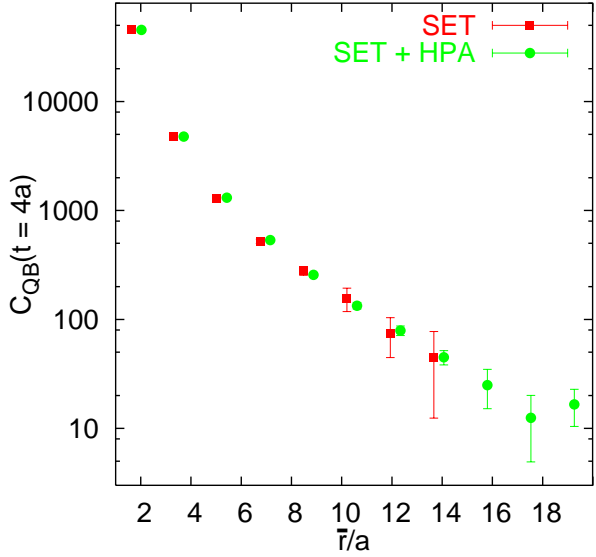}}
\caption {The effect of HPA on SET for the example of $C_{QB}$ at $t=4a$, as a function of the distance $\overline{r}$. The normalization is arbitrary and
we have employed local sources and sinks.}
\label{fig:compare4}
\end{figure}

In Figure~\ref{fig:compare4} we show the local-local $C_{QB}$ matrix element
at fixed $t=4a$ as a function of $\overline{r}$, both for SET alone and
for HPA SET.
All distances are along the spatial diagonal, i.e.\ when increasing
$\overline{r}$ by $\approx \sqrt{3}a$, the exponent $k$ increases by
three units. At our largest separation
$\overline{r}/a\approx 11\sqrt{3}\approx
19.05$ we have $k=33$! The error reduction factors
turn out to be fairly time-independent.
At the largest distance accessible without HPA $\overline{r}/a\approx
8\sqrt{3}\approx 13.9$, the error reduction
is almost five-fold. Potentially, this
should be even more impressive for
$C_{BB}^{\rm con}$, within which
two spatial light quark propagators appear. Unfortunately,
the method cannot be combined with our present smearing function, which is
essential for accessing the physical ground states.

\subsection{Notation and spectral decomposition}
\label{sec:theo}
In order to set the stage for the interpretation of our numerical data,
we detail in this section our notations
in connection with the spectral decompositions for
the different matrix elements,
Eq.~(\ref{eq:fullcorr}). 
We will assume an infinite
extent in time direction  of the lattice 
and asymptotic behaviour of all correlators. We
often suppress the
distance dependence, ${\mathbf r}$, from the expressions.

We define $B_i\overline{B}_j$ pair creation operators
${\mathcal B}_{ij}$. These create a light antiquark of flavour $j$ and a
quark of flavour $i$, besides the static sources.
The states created by ${\mathcal B}_{ij}$, $i\neq j$,
constitute a subset of the flavour non-singlet (``$I=1$'') states.
The remaining $n_f-1$ $I=1$ states are given by traceless
linear combinations of the ${\mathcal B}_{ij}$ diagonal elements.
In addition, a flavour singlet ($I=0$)
creation operator, ${\mathcal B}_s=\frac{1}{\sqrt{n_f}}
\sum_{i=1}^{n_f}{\mathcal B}_{ii}$,
can be constructed.
${\mathcal B_a}$ represents one of the $n_f^2-1$ members of the class
of $I=1$ operators.

Flavour singlet states are created both, by ${\mathcal B}_s$
and by ${\mathcal Q}$, the operator that only contains
static quark-antiquark spinors. We remind the
reader of the definitions Eq.~(\ref{eq:norm}) and
Eq.~(\ref{eq:mixi}),
$|Q\rangle={\mathcal Q}|0\rangle$,
$|B\rangle= {\mathcal B}_s|0\rangle$,
with vacuum state $|0\rangle$.
Obviously, $\langle Q|0\rangle=\langle B|0\rangle=0$.
We denote the orthonormal eigenstates
in the flavour singlet sector as
$|n\rangle$, $n=1,2,\ldots$, with energies
$E_{n+1}\geq E_n$.

For the labelling of the $I=1$ sector, we follow the convention of
Eq.~(\ref{eq:mixii}) and define, $|B_a\rangle={\mathcal B}_a|0\rangle$.
All flavour non-singlet
eigenstates share the same energy
spectrum, $E_n^a$. 
We label eigenstates of the
$I=1$ sector, within the class of states with energy $E_n^a$,
by $|n_a\rangle$, $n=1,2,\ldots$.

Just for annotation at this point: while the $I=1$ states
decouple from
the $I=0$ states there will still be mixing between $I=1$ states and
two meson states, containing for instance
a $\overline{Q}Q$ plus a $\pi$. A
calculation of the $I=1$ correlation matrix,
analogous to the $I=0$ sector,
is beyond the scope of the present paper.

Note that we smear the $B_{\mathbf r}\overline{B}_{\mathbf 0}$ creation
operator at position ${\mathbf 0}$ while the sink is smeared at ${\mathbf r}$.
This means that the creation operator
${\mathcal B}_{s|a}^{i}=\overline{Q}_{{\mathbf r}}
\gamma_5q_{{\mathbf r}}\bar{q}_{{\mathbf 0}}\Phi \gamma_5Q_{{\mathbf 0}}$
within Eq.~(\ref{eq:wil2}), where the smearing function $\Phi$ acts
on the quark at position ${\mathbf 0}$, is not the Hermitian adjoint of the 
annihilation operator,
${\mathcal B}_{s|a}^{f}\neq {\mathcal B}_{s|a}^{i\dagger}$. The subscripts
``$i$'' and ``$f$'' stand for initial and final,
respectively. Hence, strictly speaking, one has to distinguish
between $\langle B^f_{(a)}|$
and $|B^i_{(a)}\rangle$.
From a practical point of view, the highly
satisfying ground state
dominance of our data renders this distinction obsolete.

We can decompose,
\begin{eqnarray}\label{eq:deco1}
|Q\rangle = \sum_{n>1}\langle n|Q\rangle|n\rangle
=\sum_{n}Q_n|n\rangle,\\
|B^i\rangle = \sum_{n>1}\langle n|B^i\rangle|n\rangle
=\sum_{n}B^i_n|n\rangle,\label{eq:deco2}
\end{eqnarray}
where
\begin{eqnarray}
\sum_n |Q_n|^2&=&\langle Q|Q\rangle,\\
\sum_n {B^f_n}^*B^i_n
&=&\langle B^f|B^i\rangle,\\
\sum_n Q_n^*B^i_n&=&
\sum_n {B^f_n}^*Q_n=0.
\end{eqnarray}
Using these notations,
the matrix elements Eq.~(\ref{eq:fullcorr})
read (neglecting the overall energy
off-set $2m_Q$),
\begin{eqnarray}
C_{QQ}(t)&=&\www=\sum_{n}
|Q_n|^2e^{-E_nt},\label{eq:c11sp}\\
C_{QB}(t)&=&\sqrt{n_f}\wwb=
\sum_{n}\mbox{Re}(Q_n^*B^i_n) e^{-E_nt},\\
C_{BB}(t)&=&\wbbd-n_f\wbbc=
\sum_{n}\mbox{Re}({B^f_n}^*B^i_n) e^{-E_nt},\label{eq:c12sp}\\
C_{BB}^{\rm dis}(t)&=&\wbbd=\sum_{n}\mbox{Re}({B^{f*}_{a,n}}B^i_{a,n})
e^{-E_n^at}.
\label{eq:c22dsp}
\end{eqnarray}
Note that $\mbox{Re}(Q_n^*B^i_n)=\mbox{Re}(Q_n^*B^f_n)$.

The normalization of our correlation matrix is such that
$C_{\alpha\beta}(t)>0$ for $t\rightarrow\infty$, i.e.\ the ground state
amplitudes are always positive. However, within $C_{QB}$ and $C_{BB}$,
excited state amplitudes can be negative. 

For $r<r_c$ the ground state
$|1\rangle$ will
be dominated by a $|Q\rangle$-type
component, whereas the first excitation $|2\rangle$ has a large $|B\rangle$
contribution\footnote{It turns out that (hybrid) excitations
of $|Q\rangle$-nature are energetically higher than $E_2$.}.
For $r>r_c$ this correspondence will interchange.
We view  a signal $C_{QB}\neq 0$ as an ``explicit''
signature for mixing while
a verification of an
$E_1$ signal in $C_{QQ}(t)$  at $r>r_c$ (string decay) or
within $C_{BB}(t)$ at $r<r_c$ will
be referred to as an ``implicit'' mixing effect.

\section{Investigation of individual matrix elements}
\label{sec:simu}
We set the stage for the investigation of the mixing problem
by comparing individual matrix elements to theoretical
expectations. We first discuss $C_{BB}$ and then
combinations of different components of the correlation matrix,
before we attempt to detect implicit mixing effects.
\subsection{Large time asymptotics}
We remark that $C_{BB}(t)$ contains a
disconnected and a connected contribution.
The disconnected term coincides with the $I=1$ diagram
Eq.~(\ref{eq:c22dsp}),
and the states it couples to are orthogonal  to
the $I=0$ sector: any implicit mixing can only be mediated through
$C_{BB}^{\rm con}$.
This means that at $0<r<r_c$,
\begin{equation}
\label{eq:abc}
C_{BB}(t)\rightarrow -C_{BB}^{\rm con}(t)\quad(t\rightarrow\infty):
\end{equation}
the connected diagram will dominate at asymptotically large $t$.
This is in contrast to the situation at small to moderately large times,
where the overlaps
$|B_1|\ll |B_2|$ [cf.\
Eqs.~(\ref{eq:deco2}) and (\ref{eq:c12sp})] warrant
a disconnected diagram dominance.
We shall see below that Eq.~(\ref{eq:abc}) in fact turns out to be
valid for any
$r>0$, including $r>r_c$.

\begin{figure}[th]
\hspace{-0.1in}
\epsfxsize=0.9\columnwidth
\centerline{
\epsffile{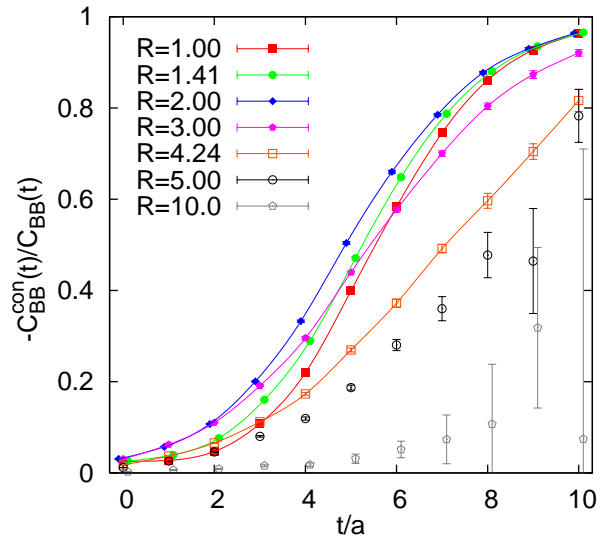}}
\caption {The ratio $-C_{BB}^{\rm con}/C_{BB}$ as a function of $t$ for various
$r=Ra$. The splines are drawn to guide the eye.}
\label{fig:rat1}
\end{figure}
\begin{figure}[th]
\hspace{-0.1in}
\epsfxsize=0.9\columnwidth
\centerline{
\epsffile{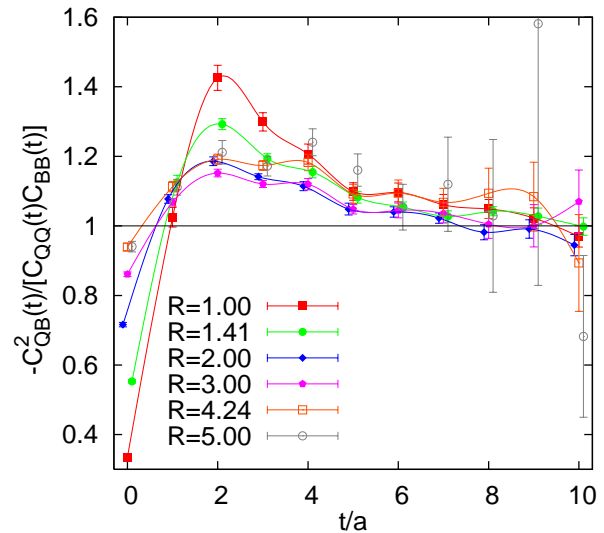}}
\caption {The ratio $-C_{QB}^2/(C_{QQ}C_{BB})$ as a function of $t$ for various
$r=Ra$. The splines are drawn to guide the eye.}
\label{fig:rat2}
\end{figure}

We investigate the above expectation
 in Figure~\ref{fig:rat1}, where we plot ratios
$-C_{BB}^{\rm con}(t)/C_{BB}(t)$ as functions of $t$ for a few $R=r/a$
values. Note that $Ra=6a\approx r_0\approx 0.5$~fm.
As expected, at large $t$, this ratio approaches one.
Note that $r=0$ represents a special case. In this limit, the $\overline{Q}Q$
and the $B\overline{B}$ sectors decouple and Eq.~(\ref{eq:abc}) does not
hold. As a consequence, the asymptotic limit is reached
faster for $r= 2a$ than for $r=a$ which is adjacent to $r=0$.
For $r>2a$, the speed of convergence decreases again:
the gap between the two energy levels $E_1$ and $E_2$
reduces as a function of $r$ and hence the limit is approached
less and less rapidly in $t$. At large $r$
the signal vanishes in noise, before it can approach unity.

\begin{figure}[th]
\hspace{-0.1in}
\epsfxsize=0.9\columnwidth
\centerline{
\epsffile{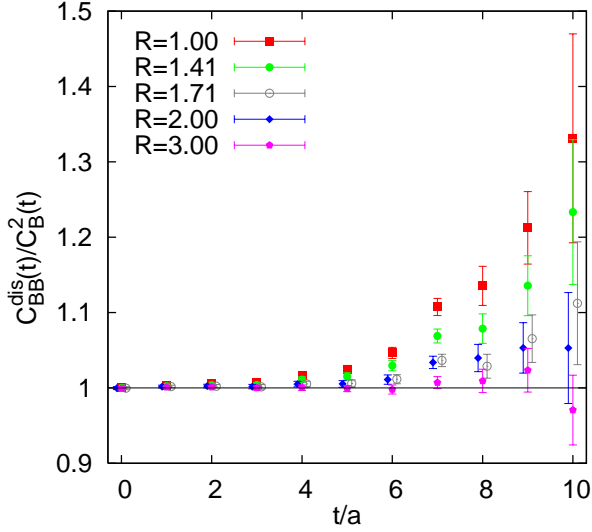}}
\caption {The ratio $C_{BB}^{\rm dis}/C_B^2$ as a function of $t$ for small
$r=Ra$.}
\label{fig:rat3}
\end{figure}

From
Eqs.~(\ref{eq:c11sp}) -- (\ref{eq:c12sp}) it is expected
that at asymptotically large times, for all $r>0$,
\begin{equation}
|C_{QB}(t)|^2\rightarrow C_{QQ}(t)C_{BB}(t)\quad(t\rightarrow\infty).
\label{eq:asyc12}
\end{equation}
In Figure~\ref{fig:rat2} we verify this expectation for some selected
distances.
The disconnected contribution to
$C_{BB}$ has to decay faster than the connected contribution. Otherwise
the corresponding
$n_f$, colour and Dirac factors would be incompatible with
Eq.~(\ref{eq:asyc12}).
Hence Eq.~(\ref{eq:abc}) above is not only valid at
$r<r_c$ but for any distance $r>0$.

In particular this means that at large $t$,
\begin{equation}
C_{QB}^2(t)\rightarrow -C_{QQ}(t)C_{BB}^{\rm con}(t),
\label{eq:asyc122}
\end{equation}
or, diagrammatically,
\begin{equation}
\wwb\times\wbw\rightarrow -\www\times\wbbc\quad,
\end{equation}
again for any $r>0$. This limiting behaviour is approached faster in time
than Eq.~(\ref{eq:asyc12}) above.

Both sides of Eq.~(\ref{eq:asyc12}) are dominated by the ground state
contribution $|1\rangle$, which results in an exponential decay
$\propto e^{-2E_1t}$. At $r<r_c$, $C_{QQ}$ couples
more strongly to this term than $C_{BB}$. This interchanges
at $r>r_c$, where the ground state is dominantly contained within
the $C_{BB}$ sector. The decoupling of  $C_{QB}$ from the
$I=1$ sector implies, $E_1\leq E_{1}^a$:
the $I=0$ ground state energy cannot be larger than the
lowest $I=1$ energy level, at any distance $r>0$.

Finally, we compare $C_{BB}^{\rm dis}(t)$ to the static-light
correlation function $C_B(t)$. If the $B$-mesons at
positions ${\bf 0}$ and ${\bf r}$ did not interact with each other
then the ratio $C_{BB}^{\rm dis}(t)/C_B^2(t)$ would be unity.
The $I=1$ $B\overline{B}$ state would merely act like the sum of two
isolated $B$
mesons. We investigate this ratio in Figure~\ref{fig:rat3} and find this
scenario to be valid
within our statistical resolution for $r\geq 2\sqrt{2}a\approx 0.23$~fm.
The increase of this ratio at large $t$ for small
$r<2\sqrt{2}a$
can be attributed to an increased overlap of the creation operator
with the $\overline{Q}Q\pi$ $I=1$ ground state, see also
Sec.~\ref{sec:shortd}.

\begin{figure}[th]
\hspace{-0.1in}
\epsfxsize=0.8\columnwidth
\centerline{
\epsffile{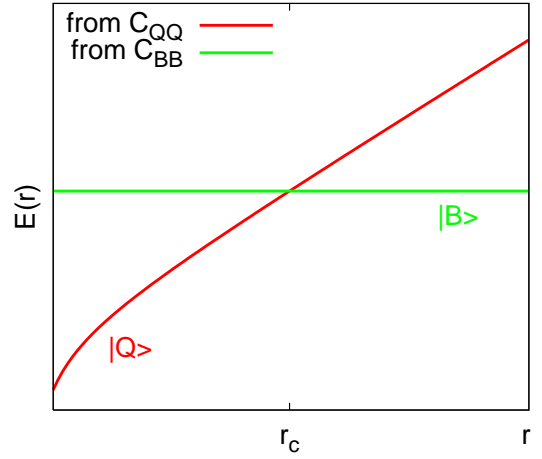}}
\caption {Expected spectrum in the absence of mixing effects. The ground state
at small $r$ has no overlap with the $|B\rangle$ sector but is only contained
within $C_{QQ}$. At $r>r_c$ the ground state is only visible in $C_{BB}$.}
\label{fig:sketch}
\end{figure}
\begin{figure}[th]
\hspace{-0.1in}
\epsfxsize=0.9\columnwidth
\centerline{
\epsffile{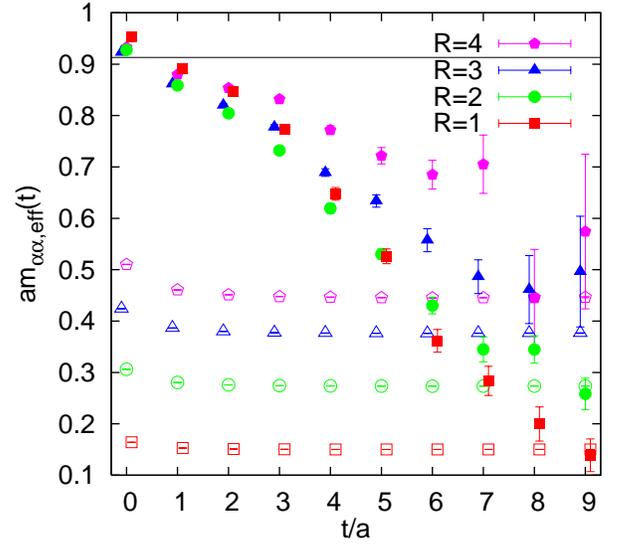}}
\caption {Effective masses. The solid symbols are from $C_{BB}(t)$,
the open symbols from the Wilson loop $C_{QQ}(t)$. The horizontal line
corresponds to twice the mass of a static-light meson.}
\label{fig:shortr}
\end{figure}
\subsection{Implicit detection of mixing effects}
We investigate whether the actual ground state energy level
is visible in $C_{BB}(t)$ at $r<r_c$ or in $C_{QQ}(t)$ at $r>r_c$.
For this purpose, we study the large $t$ behaviour of
effective masses, Eq.~(\ref{eq:effmass}). The qualitative expectation
in the absence of any mixing is sketched in Figure~\ref{fig:sketch}:
at $r<r_c$ the ground state can only be detected within $C_{QQ}$, while
at $r>r_c$ the ground state energy is given by the large $t$ behaviour
of $C_{BB}$ and will not be visible from $C_{QQ}$. Implicit mixing means that
the $C_{QQ}$ and $C_{BB}$ effective masses share the same ground state.
At $r<r_c$ the $C_{QQ}$ effective mass is expected to plateau at smaller
$t$ values than the $C_{BB}$ effective mass. At $r>r_c$ the ground state
then will become dominated by the $|B\rangle$ contribution and hardly
be visible in $C_{QQ}$.

\begin{figure}[th]
\hspace{-0.1in}
\epsfxsize=0.9\columnwidth
\centerline{
\epsffile{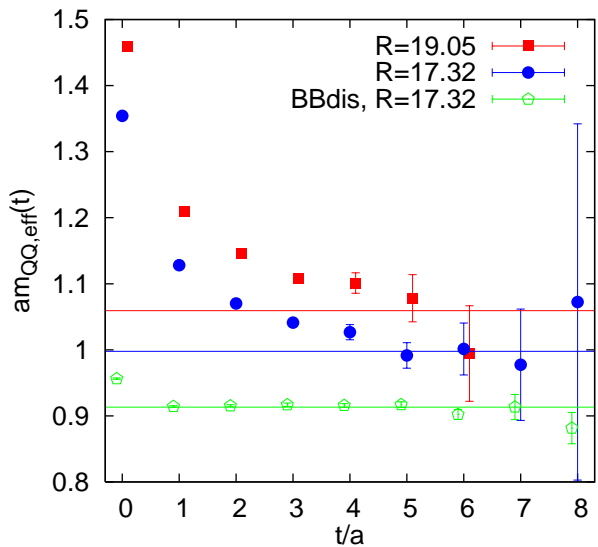}}
\caption {Effective masses from the Wilson loop $C_{QQ}$.
The lowest horizontal line is
$2m_B$. The other two horizontal lines denote the central values
of the respective excited state expectations from a global linear
plus Coulomb fit to the potential from $2<R\leq 13$ Wilson loop data.
The open pentagons, from 
$C_{BB}^{\rm dis}$, demonstrate the plateau quality within the
$C_{BB}$ sector.}
\label{fig:larger}
\end{figure}
We display the situation for small $r$ in Figure~\ref{fig:shortr}.
The open symbols, calculated from Wilson loops $C_{QQ}$,
exhibit good and early plateaus. The solid symbols, which correspond to
the matrix element $C_{BB}$, start out from
values, similar to the mass of two static-light mesons, but then 
decay towards the respective lower lying states, clearly
signalling implicit mixing effects! Note that as $C_{BB}^{\rm dis}$ alone
only projects onto the $I=1$ sector, this effect is entirely due to
the $C_{BB}^{\rm con}$ contribution, see Figure~\ref{fig:rat1}.

We are also tempted to verify implicit string breaking at large $r$.
To this end, in Figure~\ref{fig:larger} we examine effective Wilson loop
masses for two distances $r>r_c\approx 15a$: $r/a=10\sqrt{3}\approx 17.32$ and
$r/a=11\sqrt{3}\approx 19.05$. The upper two horizontal lines denote the
respective plateau value expectations
from a global linear plus Coulomb fit to the potential as obtained from
Wilson loops, in the region without string breaking, $\overline{r}<11\,a<r_c$.
We wish to see the data to deviate
from this expectation,
towards the lowest line, that corresponds to $2m_B$.
We conclude that we see no implicit indications of string breaking
in the Wilson loop data.

We also demonstrate the quality of the effective mass
plateau within the $C_{BB}$ sector
in Figure~\ref{fig:larger}
(open pentagons). Here, in the interest of small error bars,
we neglected the $C_{BB}^{\rm con}$ correction to the masses.
As expected, the $C_{BB}^{\rm dis}$ effective mass data
agree with twice the static-light mass.
We remark that we also find $E_1\approx E_1^a$ within errors, in agreement
with the level ordering
expectation, $E_{1}\leq E_{1}^a$.

The absence of an indication of implicit string breaking at large $r$
is no surprise since in a study of adjoint potentials in
$2+1$ dimensional $SU(2)$ gauge theory~\cite{Kratochvila:2003zj}, this was
only seen at physical times much bigger than ours.
We shall address the question, at what $t$ values implicit string breaking
should
become visible, in Sec.~\ref{sec:transition} below.
We will see in Sec.~\ref{sec:mixan}, where we also study explicit mixing
effects, that mixing is indeed much smaller at $r>r_c$ than at
$r<r_c$.

\section{Results}
\label{sec:result}
We present our analysis and results on the $I=0$ mixing angle and
energy levels. We then discuss string breaking as well as transition
rates, before we address the short distance behaviour of the energy levels,
both within the $I=0$ and the $I=1$ sectors. As we only work at a fixed value
of the lattice spacing,
it is convenient to display all results in this section
in lattice units, $a\approx
0.083$~fm, i.e.\ $a^{-1}\approx 2.37$~GeV.

\subsection{The mixing analysis}
\label{sec:mixan}
Our creation and annihilation operators are highly optimized,
such that the overlaps
$Q_n=\langle n|Q\rangle$ and $B_n=\langle n|B\rangle$ are close to {\em
zero} for $n\geq 3$. Hence, we base our analysis on the
simplified mixing scenario,
\begin{eqnarray}
\label{eq:mmx1}
|Q\rangle = a_Q\left(\cos\theta |1\rangle - \sin\theta |2\rangle\right),\\
|B\rangle = a_B\left(\sin\theta |1\rangle + \cos\theta |2\rangle\right),
\label{eq:mmx2}
\end{eqnarray}
truncating Eqs.~(\ref{eq:deco1}) and (\ref{eq:deco2}) after $n=2$.
In what follows, we abbreviate, $c_{\theta}=\cos\theta$ and
$s_{\theta}=\sin\theta$.
The identification, $Q_1 = a_Q c_{\theta}$, $Q_2=-a_Q s_{\theta}$,
$B_1=a_Bs_{\theta}$ and $B_2=a_Bc_{\theta}$ above guarantees
that $\langle Q|B\rangle=0$, as well as positivity of all
correlation matrix elements for large Euclidean times.

The ansatz Eqs.~(\ref{eq:mmx1}) -- (\ref{eq:mmx2}) implies that,
\begin{eqnarray}
\label{eq:mix2}
C_{QQ}(t)&=&a_Q^2\left[c_{\theta}^2\exp\left(-E_1t\right)
+s_{\theta}^2\exp\left(-E_{2}t\right)\right],\\
C_{BB}(t)&=&a_B^2\left[s_{\theta}^2\exp\left(-E_1t\right)
+c_{\theta}^2\exp\left(-E_{2}t\right)\right],\\
C_{QB}(t)&=&a_Qa_Bs_{\theta}c_{\theta}\left[\exp\left(-E_1t\right)-
\exp\left(-E_{2}t\right)\right].
\label{eq:mix3}
\end{eqnarray}
We also determine the correlation functions at $t=0$. This enables us
to implement
the normalization, $C_{QQ}(0)=C_{BB}(0)=1$. In this case, the
$a_Q$ and $a_B$ values, fitted at large $t$,
can be interpreted as the overlaps
of our respective trial wave functions
with the $n\leq 2$ eigenstate sector, with optimal
value, $a_Q^2=a_B^2=1$. While $a_Q^2\leq 1$, this is not necessarily
so for $a_B^2$ as here additional exponentials can come in with negative
weight.

We attempted to model corrections to Eqs.~(\ref{eq:mix2}) --
(\ref{eq:mix3}), by adding 
additional exponentials to our fits.
The overlaps
$\langle n|B\rangle$ for $n\geq 3$
were so tiny that, except
at $r\leq 2a$, we were unable to
detect any additional masses in the $C_{BB}$ channel.
It was however possible to add an additional
excitation to the $|Q\rangle$ channel. This exponential
then also couples to the
$C_{QB}$ element. Results of such eight parameter fits were very
compatible with those of the simultaneous five parameter fits
introduced above,
with parameters $\theta, a_Q, a_B, E_1$ and $E_2$.
However, these eight parameter fits were not stable at all
distances. In contrast, the five parameter fits turned out to be very robust,
such that the results presented here are based on the parametrization
Eqs.~(\ref{eq:mix2})--(\ref{eq:mix3}). We also attempted six parameter
fits, allowing $\theta$ to take two different values within
Eq.~(\ref{eq:mmx1}) and Eq.~(\ref{eq:mmx2}). This however did not
improve the $\chi^2/N_{DF}$ qualities and the two $\theta$-angles
turned out to agree within errors. We conclude that the mixing scenario
Eqs.~(\ref{eq:mix2})--(\ref{eq:mix3}) is preferred by the data.

For each $r$ we
carefully checked the quality of the fits and
the stability of the parameter values with respect to
variations of the fit range. For each of the three matrix elements
we determined a $t_{\alpha\beta,\min}$
value. $a_B$ turned out to be much closer to
unity than $a_Q$ and we found,
$t_{QQ,\min}\geq t_{QB,\min}\geq t_{BB,\min}$.
The fit ranges that we employed in our final
analysis are,
\begin{widetext}
\begin{eqnarray} 
\label{eq:range1}
t_{QQ,\min}=4a,\, t_{QB,\min}=3a,\, t_{BB,\min}=2a
&&\mbox{for}\quad \overline{r}>15.5a,\\
t_{QQ,\min}=t_{QB,\min}=4a,\, t_{BB,\min}=2a &&\mbox{for}\quad
15.5a\geq\overline{r}>13a,\\ 
t_{QQ,\min}=5a,\, t_{QB,\min}=4a,\, t_{BB,\min}=2a &&\mbox{for}\quad
13a\geq\overline{r}>12a,\\
t_{QQ,\min}=6a,\, t_{QB,\min}=5a,\, t_{BB,\min}=2a &&\mbox{for}\quad
12a\geq\overline{r}>8a,\\
t_{QQ,\min}=6a,\, t_{QB,\min}=5a,\, t_{BB,\min}=3a &&\mbox{for}\quad
\overline{r}\leq 8a.
\label{eq:range5}
\end{eqnarray}
\end{widetext}

\begin{table*}[h]
\caption{\label{tab:fit}The energy levels $E_1$ and $E_2$, as well as
the mixing angle $\theta$, the
overlaps $a_Q$ and $a_B$ and the transition rate $g$, defined in Eq.~(\protect\ref{eq:tran}). Note that at $r=0$, $E_1=0$ and
the ground state energy is, $a(E_g-2m_B)=-0.59(17)$. $E_2$ denotes the
first excitation above this state.}
\begin{ruledtabular}
\begin{tabular}{c|c|c|c|c|c|c|c}
$\overline{r}/a$&$a[E_1-2m_B]$&$a[E_2-2m_B]$&$a[E_2-E_1]$&$\theta$&$a_Q$&$a_B$&$ag$\\\hline
$r=0$ &-0.913 (3)&0.021 (7)&0.934 (7)& 0       & 1       &0.978 (3)& 0\\
 1.365&-0.759 (4)&0.021 (7)&0.780 (8)&0.129 (2)&1.016(10)&0.985(11)&0.0996 (5)\\
 1.442&-0.706 (4)&0.028 (5)&0.734 (7)&0.168 (2)&1.019(10)&0.995 (9)&0.1210 (5)\\
 1.826&-0.648 (4)&0.035 (4)&0.683 (6)&0.196 (3)&1.022(10)&1.003 (7)&0.1304 (7)\\
 1.855&-0.634 (4)&0.032 (5)&0.667 (7)&0.212 (3)&1.022(10)&1.000 (8)&0.1369 (6)\\
 2.836&-0.539 (4)&0.036 (4)&0.575 (5)&0.239 (3)&1.025(11)&1.007 (6)&0.1323 (6)\\
 2.889&-0.529 (4)&0.034 (4)&0.564 (5)&0.246 (3)&1.026(10)&1.005 (7)&0.1330 (6)\\
 3.513&-0.489 (4)&0.033 (4)&0.522 (5)&0.239 (3)&1.021(10)&1.005 (6)&0.1199 (7)\\
 3.922&-0.460 (4)&0.029 (4)&0.489 (5)&0.231 (3)&1.010(10)&1.002 (6)&0.1089 (6)\\
 4.252&-0.444 (4)&0.030 (3)&0.474 (5)&0.221 (3)&1.001(11)&1.004 (5)&0.1014 (6)\\
 4.942&-0.404 (3)&0.026 (4)&0.430 (4)&0.203 (2)&0.989 (9)&1.003 (5)&0.0849 (5)\\
 5.229&-0.392 (3)&0.026 (4)&0.418 (4)&0.193 (2)&0.983 (9)&1.004 (6)&0.0787 (6)\\
 5.666&-0.370 (4)&0.020 (3)&0.389 (5)&0.184 (3)&0.976(11)&0.996 (5)&0.0700 (4)\\
 5.954&-0.360 (3)&0.019 (4)&0.379 (4)&0.177 (3)&0.955 (9)&0.997 (5)&0.0657 (6)\\
 6.953&-0.315 (4)&0.016 (3)&0.331 (4)&0.163 (3)&0.941(12)&0.995 (5)&0.0531 (6)\\
 6.962&-0.315 (3)&0.017 (4)&0.332 (4)&0.166 (3)&0.937 (8)&0.996 (6)&0.0542 (7)\\
 7.079&-0.307 (4)&0.010 (3)&0.318 (4)&0.169 (3)&0.944(10)&0.984 (4)&0.0527 (5)\\
 7.967&-0.273 (3)&0.010 (4)&0.283 (5)&0.170 (4)&0.917 (8)&0.986 (7)&0.0471 (6)\\
 8.492&-0.252 (5)&0.008 (2)&0.260 (5)&0.172 (4)&0.907(15)&0.982 (2)&0.0439 (6)\\
 8.680&-0.239 (5)&0.005 (2)&0.244 (5)&0.175 (5)&0.920(15)&0.978 (2)&0.0418 (8)\\
 8.971&-0.222 (5)&0.006 (2)&0.228 (5)&0.181 (4)&0.926(13)&0.980 (2)&0.0405 (7)\\
 9.905&-0.197 (7)&0.007 (2)&0.204 (7)&0.180 (5)&0.875(17)&0.982 (2)&0.0358 (8)\\
 9.974&-0.187 (6)&0.006 (2)&0.193 (5)&0.186 (6)&0.891(17)&0.980 (2)&0.0351 (8)\\
10.408&-0.182 (9)&0.005 (2)&0.187 (9)&0.179(10)&0.849(24)&0.979 (2)&0.0327(10)\\
10.977&-0.145(11)&0.002 (2)&0.147(10)&0.202(16)&0.878(31)&0.977 (2)&0.0289 (8)\\
11.319&-0.145 (8)&0.006 (2)&0.150 (8)&0.191(11)&0.837(20)&0.980 (2)&0.0280 (8)\\
12.138&-0.119 (9)&0.005 (2)&0.125(10)&0.194(15)&0.811(23)&0.980 (2)&0.0235 (6)\\
12.733&-0.092 (7)&0.006 (1)&0.097 (7)&0.227(16)&0.814(15)&0.980 (2)&0.0214 (5)\\
13.869&-0.036 (6)&0.008 (2)&0.044 (5)&0.384(58)&0.823(14)&0.981 (2)&0.0151 (6)\\
14.147&-0.039 (5)&0.007 (2)&0.046 (5)&0.345(39)&0.792 (9)&0.981 (2)&0.0147 (5)\\
14.288&-0.032 (5)&0.007 (2)&0.039 (4)&0.395(49)&0.794(12)&0.981 (3)&0.0140 (6)\\
14.463&-0.026 (5)&0.007 (2)&0.033 (3)&0.498(71)&0.793(12)&0.978 (3)&0.0137 (6)\\
14.605&-0.016 (4)&0.009 (3)&0.025 (3)&0.59 (11)&0.801(12)&0.981 (2)&0.0115 (6)\\
14.704&-0.017 (4)&0.009 (3)&0.027 (2)&0.69 (11)&0.794(11)&0.977 (3)&0.0131 (5)\\
15.008&-0.011 (3)&0.011 (3)&0.022 (1)&0.87 (12)&0.784(11)&0.977 (2)&0.0110 (5)\\
15.176&-0.005 (3)&0.018 (5)&0.022 (3)&1.01 (12)&0.787(12)&0.981 (2)&0.0100 (7)\\
15.372&-0.003 (2)&0.027 (7)&0.030 (5)&1.204(78)&0.792(14)&0.980 (3)&0.0100 (6)\\
15.561&-0.002 (2)&0.027 (6)&0.029 (5)&1.15 (10)&0.771(14)&0.982 (2)&0.0109 (4)\\
15.600&-0.002 (2)&0.037 (8)&0.039 (8)&1.305(61)&0.793(15)&0.980 (2)&0.0100 (5)\\
17.331& 0.001 (2)&0.095(13)&0.094(12)&1.501(12)&0.756(21)&0.981 (2)&0.0066 (5)\\
19.063& 0.003 (2)&0.164(20)&0.161(20)&1.546 (4)&0.736(29)&0.982 (2)&0.0040 (4)
\end{tabular}
\end{ruledtabular}
\end{table*}

We determined the correlation matrix elements $C_{\alpha\beta}(t)$
for $0\leq t\leq 10$.
However, the quality
of the $C_{QQ}$ and $C_{QB}$ data only allowed us to use
$t_{\max}=9$ for $\overline{r}\geq 12$.
For the discussion of the individual matrix elements presented
in the previous sections,
we calculated jackknife and bootstrap
errors. Both showed consistent results.
Hence, in this more complicated
mixing analysis we restrict ourselves to the jackknife method.

Prior to the fits we transformed the data:
\begin{equation}
C_{\alpha\beta}(t)\rightarrow C_{\alpha\beta}(t)/C_B^2(t).
\end{equation}
This automatically normalizes the energy levels
with respect to $2m_B$, removing the self energy $2m_Q$. In principle,
this procedure bears the risk of introducing additional
energy levels that are present in the static-light correlation function
but not within the $\overline{Q}Q/B\overline{B}$ system.
Figure~\ref{fig:compare} confirms nicely that at $t>a$ no such levels are
statistically detectable. $C_B(0)$, however,
turns out to be about 2~\% larger than an exponential extrapolation
down from $t\geq a$ suggests. We correct for this in the calculation
of the overlaps $a_Q$ and
$a_B$. The results for all fit parameters are summarized
in Table~\ref{tab:fit}.

$\theta$ is the mixing angle
of the physical eigenstates $|1\rangle, |2\rangle$
with respect to the Fock
basis $|Q\rangle, |B\rangle$, used in the simulation.
We invert Eqs.~(\ref{eq:mmx1}) -- (\ref{eq:mmx2}),
adapting the normalization $|Q\rangle\mapsto a_Q^{-1}|Q\rangle$,
$|B\rangle\mapsto a_B^{-1}|B\rangle$:
\begin{eqnarray}
\label{eq:mx1}
|1\rangle&=&\cos\theta|Q\rangle+\sin\theta|B\rangle,\\
|2\rangle&=&-\sin\theta|Q\rangle+\cos\theta|B\rangle.
\label{eq:mx2}
\end{eqnarray}
In our $t_{\alpha\beta}\geq t_{\alpha\beta,\min}$ analysis
we effectively encounter this
idealized picture and truncate the
eigenbasis at $n=2$, which is supported by the data.
However, 
we also truncate the Fock basis after the
$\overline{Q}q\bar{q}Q$ sector. In general, the physical eigenstates
will receive contributions from higher Fock states too and hence
there will be a (slight) model dependence in the determination of
$\theta$.

The limit ${\mathbf r}={\mathbf 0}$ represents a special case.
In this limit,
\begin{equation}
\wbw\propto {\mathbf 0}\cdot\langle \bar{q}_0\boldsymbol{\gamma}(1-\gamma_4)q_0
\rangle_U=0,
\end{equation}
since
$\langle \bar q\gamma_i q\rangle_U=\langle \bar q\gamma_i\gamma_4 q\rangle_U
=0$: the two $I=0$ eigenstates decouple and hence $\theta=0$.
Moreover, $\langle W({\mathbf 0},t)\rangle_U
=const$, which means that the vacuum is the ground state.
$\overline{Q}$ and $Q$ annihilate: $E_1(0)-2m_B=-2m_B$. The ground state
overlap at this point is $a_Q=1$, by definition, and $C_{QB}$ is undefined.
At $r=0$ we perform a two exponential fit for $t\geq a$
to $C_{BB}(t)$, with exponents $E_g$ and $E_2$.
The lower mass (without subtracting $2m_B$)
is $E_g=0.32(17)a^{-1}$ but with tiny overlap:
$a_g^2=0.0045(15)$. This mass should coincide with
the mass of two interacting pions or with the scalar $f_0$ mass,
see also Sec.~\ref{sec:shortd} below.
We have $2m_{\pi}=0.55(1)a^{-1}$ for two non-interacting pions while
the mass of the lightest scalar
is~\cite{Bali:2000vr} $m_{0^{++}}=0.71(6)a^{-1}$. Both values are
compatible within two standard deviations with the fitted $E_g$ value above.

\begin{figure}[th]
\hspace{-0.1in}
\epsfxsize=0.9\columnwidth
\centerline{
\epsffile{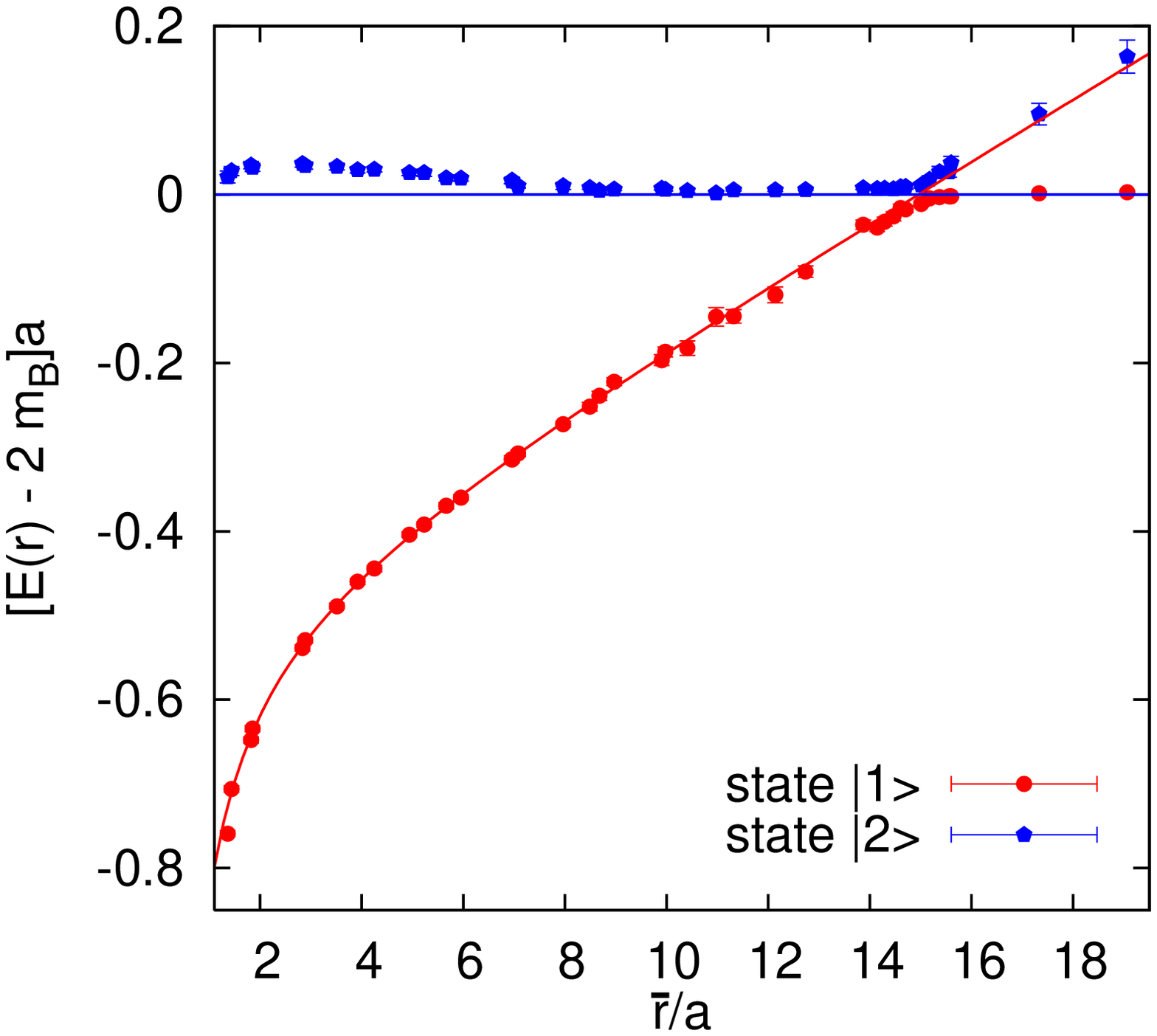}}
\caption {The two energy levels, as a function of $\overline{r}$,
normalized with respect to $2m_B$ (horizontal line). The curve
corresponds to the three parameter fit to $E_1(\overline{r})$,
Eqs.~(\protect\ref{eq:funnel})--(\protect\ref{eq:funnel3}), for
$0.2\,\mbox{fm}\leq\overline{r}\leq 0.9\,\mbox{fm}<r_c$.}
\label{fig:break1}
\end{figure}
\begin{figure}[ht]
\hspace{-0.1in}
\epsfxsize=0.9\columnwidth
\centerline{
\epsffile{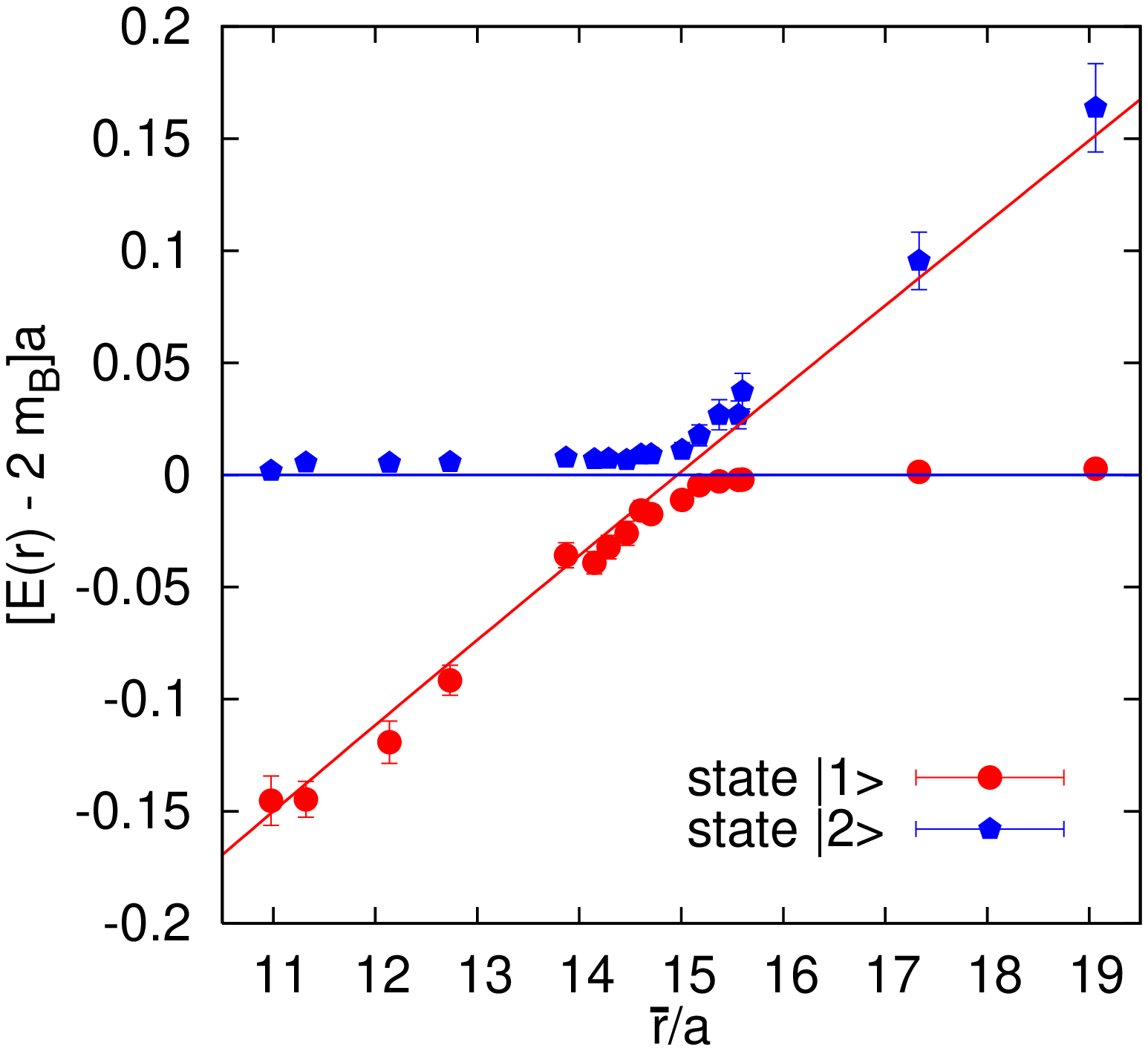}}
\caption {The same as Figure~\protect\ref{fig:break1}, for the string breaking
region.}
\label{fig:region}
\end{figure}
\subsection{String breaking}
In Figure~\ref{fig:break1} we plot the two energy levels, normalized
with respect to $2m_B$, as a function of the $O(\alpha_s)$ improved
lattice distance $\overline{r}$, Eq.~(\ref{eq:rbar}). Note that
string breaking
takes place at a distance $r_c\approx 15a\approx 1.25$~fm.
The implications with
respect to the $n_f=2+1$ QCD situation with realistic quark masses
are discussed in Sec.~\ref{sec:pheno} below. The curve corresponds to
the three-parameter fit,
\begin{equation}
\label{eq:funnel}
E_1(r)=V_0+\sigma r-e/r,
\end{equation}
with fit range $0.2\,\mbox{fm}\leq\overline{r}\leq 0.9$~fm.
For the normalization we find,
$V_0=2m_B-0.509(8)a^{-1}$ while string tension and Coulomb coefficient
are respectively,
\begin{eqnarray}
\label{eq:funnel2}
\sqrt{\sigma}&=&0.1888(29)\,a^{-1},\\
e&=&0.362(16).
\label{eq:funnel3}
\end{eqnarray}
The fit implies a Sommer parameter,
\begin{equation}
r_0^2\left.\frac{dE_1(r)}{dr}\right|_{r=r_0}=1.65,
\end{equation}
of
\begin{equation}
r_0=6.009(53)a\approx 0.5\,\mbox{fm},
\end{equation}
which we use to translate the lattice scale $a$ into physical units.

On the scale of Figure~\ref{fig:break1}, the energy gap
$\Delta E_c=\min_r[E_2(r)-E_1(r)]$ is barely visible.
Therefore, we enlarge the string breaking
region in Figure~\ref{fig:region}.
We define the string breaking distance as the distance
where the energy gap is minimal: $E_2(r_c)-E_1(r_c)=\Delta E_c$.

\begin{figure}[th]
\hspace{-0.1in}
\epsfxsize=0.9\columnwidth
\centerline{
\epsffile{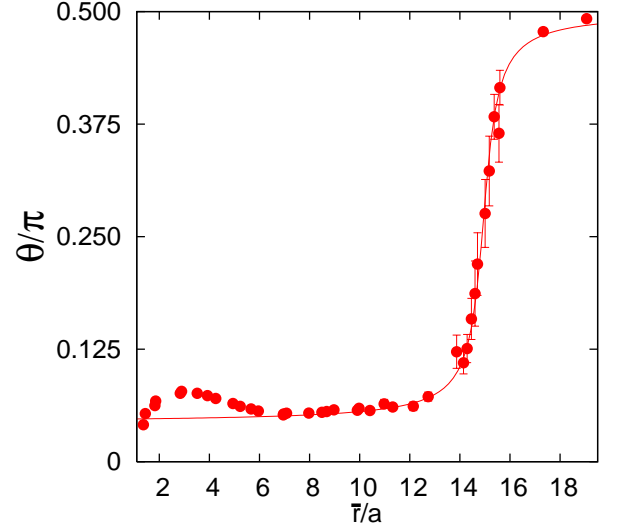}}
\caption {The mixing angle $\theta$, as a function of $\overline{r}$.
The curve corresponds to the parametrization
Eqs.~(\protect\ref{eq:thetafit})--(\protect\ref{eq:thetafit3}).}
\label{fig:angle}
\end{figure}

Not only the two energy levels play a role in the mixing dynamics
but also the mixing
angle $\theta$ of Eqs.~(\ref{eq:mx1}) and (\ref{eq:mx2}).
In Figure~\ref{fig:angle}
we depict $\theta$ as a function
of $r$. For $r<r_c$, the overlap
$Q_1$ will be larger than
$B_1$ and hence
$\theta<\pi/4$. For $r\rightarrow\infty$ the $\overline{Q}Q$ content
of the ground state will vanish and $\theta\rightarrow\pi/2$.
The Figure reveals that while this large $r$ limit is rapidly approached
for $r>r_c$, the ground state at small $r$ contains
a significant
$B\overline{B}$ admixture: for instance,
$\sin^2[\theta(8a)]\approx 0.03$. Furthermore, there is a
``bump'' at small $r$ in $\theta(r)$ as well
as in $E_2(r)$, before $\theta$ is forced to approach {\em zero}
at $r\rightarrow 0$\footnote{Note that $\overline{\mathbf{0}}\approx 0.92 a$.},
where $C_{QB}(t)=0$.
This bump is likely to be related
to light meson exchange, where in our study $m_{\pi}^{-1}\approx 4a$.

\begin{figure}[th]
\hspace{-0.1in}
\epsfxsize=0.9\columnwidth
\centerline{
\epsffile{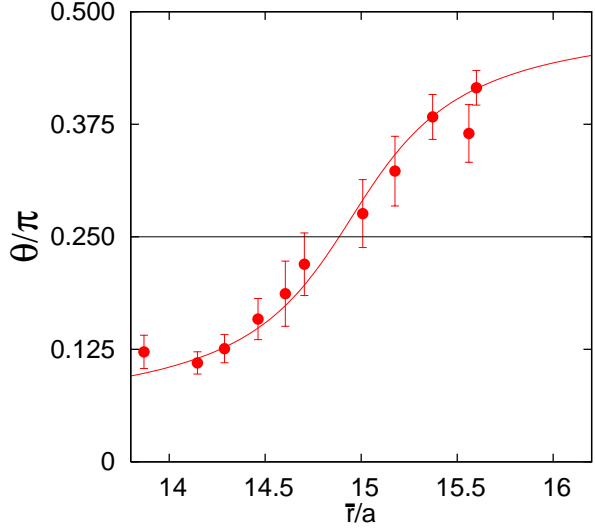}}
\caption {The same as Figure~\protect\ref{fig:angle}, for
the string breaking region.}
\label{fig:angle2}
\end{figure}
\begin{figure}[th]
\hspace{-0.1in}
\epsfxsize=0.9\columnwidth
\centerline{
\epsffile{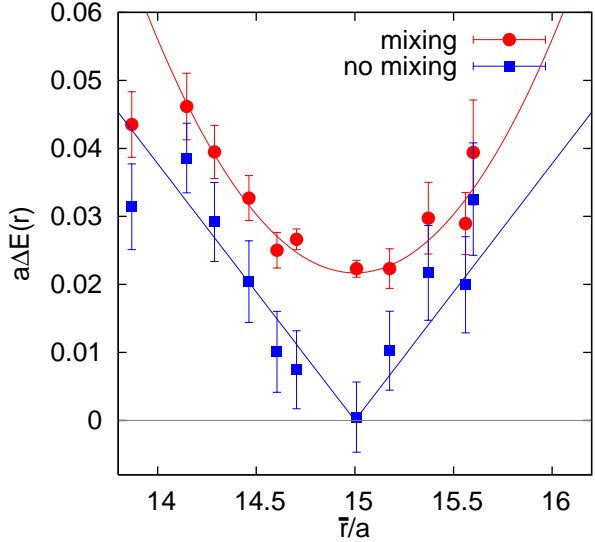}}
\caption {The energy gap $\Delta E=E_2-E_1$ with (circles) and
without (squares) mixing.}
\label{fig:gap}
\end{figure}
The curve corresponds to a phenomenological
three parameter fit to the $0.9\,\mbox{fm}\approx
11a\leq\overline{r}\leq 19a\approx 1.6\,\mbox{fm}$ data:
\begin{equation}
\label{eq:thetafit}
\theta(r)=
\frac{c}{2}\left\{\arctan\left[d(r-r_s)\right]-\frac{\pi}{2}\right\}+
\frac{\pi}{2},
\end{equation}
with parameter values,
\begin{eqnarray}
\label{eq:rc1}
r_s&=&14.95(12)a,\\
d&=&2.31(21)a^{-1},\\
c&=&0.914(6).
\label{eq:thetafit3}
\end{eqnarray}
The increase of $\theta$ with respect to $r$ for $r\approx r_s$ is given
by, $d\theta(r)/dr|_{r=r_s}=cd/2=0.34(3) \pi\,a^{-1}$.
Our distance-resolution
clearly allows us to resolve the
mixing dynamics at $r\approx r_c$. We enlarge this region
in Figure~\ref{fig:angle2}.

Finally, in Figure~\ref{fig:gap}, we investigate the difference
$\Delta E(r)=E_2(r)-E_1(r)$
in the string breaking region. The circles
represent the results from our mixing analysis while the squares are
extracted from fits to the Wilson loops $C_{QQ}$ and the $I=1$ 
$B\overline{B}$ operator $C_{BB}^{\rm dis}$
alone. This resembles the situation in the quenched
approximation  where no string breaking or mixing occurs.
We perform a quadratic
fit in the region $14a\leq \overline{r}\leq 16a$,
\begin{equation}
\Delta E(r)=\Delta E_c + b^3 (r-r_c)^2.
\end{equation}
The resulting parameter values are,
\begin{eqnarray}
r_c &=& 15.00(8)\,a,\\
\Delta E_c&=&0.0217(9)\,a^{-1},\\
b&=&0.325(14)\,a^{-1}.
\end{eqnarray}
The position of the minimal energy gap $r_c=15.00(8)a$
is in perfect agreement with the value
$r_s=14.95(12)a$ of Eq.~(\ref{eq:rc1}), at which $\theta=\pi/4$.
Translated into physical units
we obtain a minimal energy gap, $\Delta E_c\approx 51(3)$~MeV, and a string
breaking distance,
\begin{equation}
\label{eq:RC}
r_c=2.496(26)\,r_0\approx 1.248(13)\,\mbox{fm}.
\end{equation}
The
errors quoted are purely statistical and do not contain the 5~\% uncertainty
of  $r_0\approx 0.5$~fm or the deviation of $n_f=2$ and $m\lesssim m_s$
from the real QCD situation.

\begin{figure}[th]
\hspace{-0.1in}
\epsfxsize=0.9\columnwidth
\centerline{
\epsffile{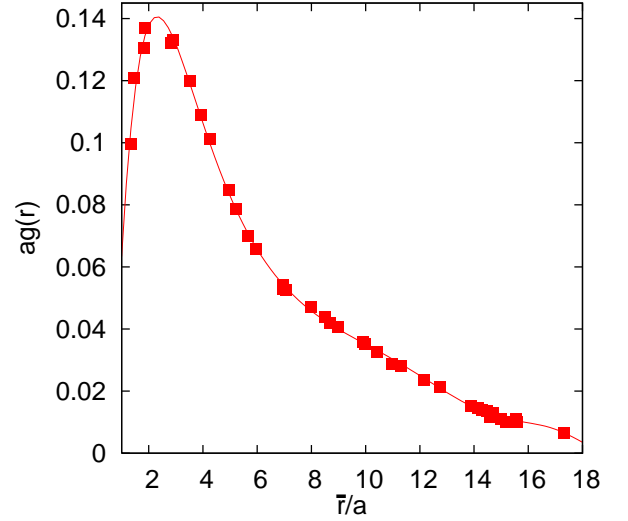}}
\caption {The transition rate $g$ between $|B\rangle$ and $|Q\rangle$
states, as a function of $\overline{r}$.}
\label{fig:decay}
\end{figure}

\subsection{Transition rates}
\label{sec:transition}
We assume that the elements of our mixing
matrix only couple to the lowest two QCD eigenstates within the
appropriate static-static
sector. In this limit, for each $r$, we encounter
a quantum mechanical two-state system. Our two test wave functions
are not QCD eigenstates and, therefore, the off-diagonal matrix
elements
$C_{QB}(t)$ assume non-trivial values.
The transition rate, governing string fission at $r>r_c$
and fusion at $r<r_c$, is given by,
\begin{equation}
\label{eq:tran}
g=\left.\frac{dC_{QB}(t)}{dt}\right|_{t=0}\frac{1}{\sqrt{C_{BB}(0)C_{QQ}(0)}}.
\end{equation}
While in Euclidean time all Fock states eventually decay into
the ground state $|1\rangle$, in Minkowski space-time,
starting from such a non-eigenstate, results
in oscillations between the $\overline{Q}Q$ and $B\overline{B}$ sectors.

Obviously, our states $|Q\rangle$ and $|B\rangle$
are somewhat polluted by $n\geq 3$ excitations as evidenced by
$a_Q\neq 1$ and $a_B\neq 1$. So we have to ``wait''
for some initial relaxation
time $t_{\min}$ to pass until this equation becomes applicable.
We can easily extract
$g$ from our five parameter fits, Eqs.~(\ref{eq:mix2}) -- (\ref{eq:mix3}),
setting $C_{QQ}(0)=a_Q^2$ and $C_{BB}(0)=a_B^2$:
\begin{equation}
g(r)=\Delta E(r)\frac{\sin[2\theta(r)]}{2}.
\end{equation}

This quantity is plotted in Figure~\ref{fig:decay} and the resulting
values are displayed
in the last column of Table~\ref{tab:fit}.
At the string breaking point, $\theta\approx \pi/4$ and hence
$g(r_c)\approx \Delta E_c/2\approx 25(2)$~MeV.
This means that $g(r)$
has a maximal value of around 320~MeV,
at a distance of about 0.2~fm.
The curve is a polynomial and drawn to guide the eye.
$g^{-1}$ can be interpreted as
the characteristic
Euclidean time scale, governing the decay of one state into the
other. For instance, at $r=2a$ we find
the (maximal) value $g^{-1}\approx 7.3a$ and indeed,
in Figure~\ref{fig:shortr}, at $t/a> 7$
the $C_{BB}$ effective masses (solid circles)  agree
with the $C_{QQ}$ level (open circles). We also find that the implicit
indications of mixing effects are
most pronounced at exactly the distance at which $g$ is largest.

For small $r$, $g$ decreases as it has to reach {\em zero} at $r=0$.
At large $r$, $\theta$ approaches $\pi/2$ quite rapidly, resulting
in small $g$ values too: for $r>r_c$ we find
$g<\Delta E_c/2\approx 0.011\,a^{-1}$. This means that detecting
the (dominantly $|B\rangle$) ground state of the system from
Wilson loop signals alone necessitates distances $t=O(100\,a)$.
Possibly, depending on the statistical accuracy, $t=50\,a$
might be sufficient to verify the decay of the Wilson loop signal
towards the ground state energy. In view of this,
it is no surprise that in Figure~\ref{fig:larger}
we have been unable to
verify such implicit string breaking at $r>r_c$ from $t\leq 9a$ data.

It is possible to calculate $g$ directly from the data,
without any fits. This will be a valuable consistency check.
For this purpose, the time derivative has to be
eliminated from Eq.~(\ref{eq:tran}).
It is straight forward
to derive the
approximate expression~(for a similar result, see e.g.\
Michael~\cite{Michael:2003vw}),
\begin{widetext}
\begin{equation}
\label{eq:michael}
g(t)\approx
Z\frac{C_{QB}(t)\sqrt{C_{QQ}(0)C_{BB}(0)}}
{\sum_{j=1}^{t/a-1}C_{QQ}(ja)C_{BB}(t-ja)+\frac{1}{2}\left[C_{QQ}(0)C_{BB}(t)+C_{QQ}(t)C_{BB}(0)\right]},\quad
Z=\frac{a\Delta E}{2\tanh(a\Delta E/2)}.
\end{equation}
\end{widetext}

\begin{figure}[th]
\hspace{-0.1in}
\epsfxsize=0.9\columnwidth
\centerline{
\epsffile{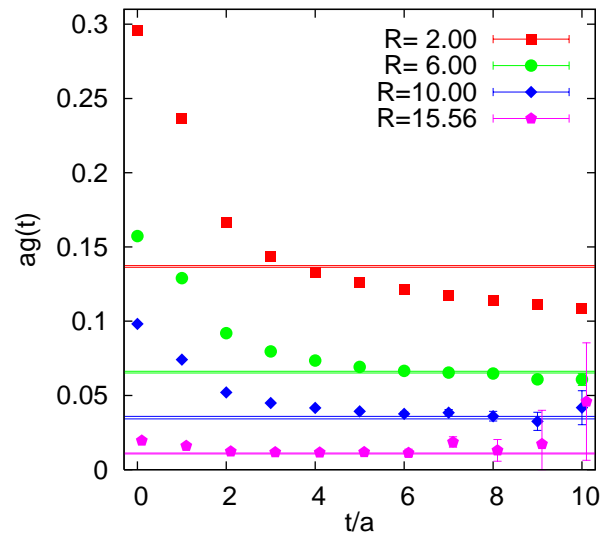}}
\caption {The modified Michael ratio Eq.~(\protect\ref{eq:michael})
for different $r=Ra$ values, as a function of $t$. The horizontal
error bands correspond to the fitted transition rates.}
\label{fig:decay2}
\end{figure}

The $Z$ term originates from
replacing a time integral by a discrete lattice sum.
The required level difference $\Delta E$
can be approximated by an effective mass, however,
for our proof-of-principle calculation we use the
$\Delta E$ values of Table~\ref{tab:fit},
extracted from our five parameter mixing fits. The largest
such correction
for the examples, displayed in
Figure~\ref{fig:decay2}, amounts to a $3.7\,\%$
upward shift of the $r=2a$ data.

If the energy gap
$\Delta E$ is large then, within the denominator,
the propagation of the lighter state
is strongly preferred over that of the heavier state
and the transition between the two states
will take place near the end points $j\approx 0$ or $j\approx t/a$.
In this case,
unless $E_3\gg E_2$,
there will be higher state contaminations and no accurate result
can be expected. If $\Delta E$ is large then
$g=s_{\theta}c_{\theta}\Delta E$ can also be large.
In the derivation of Eq.~(\ref{eq:michael}) implicit
mixing effects are neglected and due to this, at large $t$,
there will be corrections, $g=g(t)+O(gt)$.
If $g$ is sufficiently small, then
there is a chance of identifying a plateau in $g(t)$
{}from large enough (but not too large) $t$ values.

In Figure~\ref{fig:decay2}
we compare $g(t)$ approximants, obtained by use of the modified
Michael ratio
method~Eq.~(\ref{eq:michael}), with our fitted $g$ values (horizontal 
error bands).
We find good plateaus and perfect agreement with the fitted $g$ values,
except at distances $r\leq 6a$  where implicit mixing is significant
and the linear $t$ behaviour already sets in,
before the excited state contributions have died out.
In principle,
one could attempt to subtract such linear terms.

The ratio method offers a nice check of consistency.
Other than this, we see little advantage in calculating
$g(t)$ over extracting $g$ from a global fit (in $t$) of the
correlation matrix elements.
In the case of very noisy data such fits may
turn out impossible but in this case any $g(t)$ estimate will be
unreliable anyhow.
Note that within our fit ranges~Eqs.~(\ref{eq:range1})--(\ref{eq:range5}),
$g(t)$ does not show any sizeable excited state contaminations.

\subsection{Short distance $B\overline{B}$ forces}
\label{sec:shortd}
\begin{figure}[th]
\hspace{-0.1in}
\epsfxsize=0.9\columnwidth
\centerline{
\epsffile{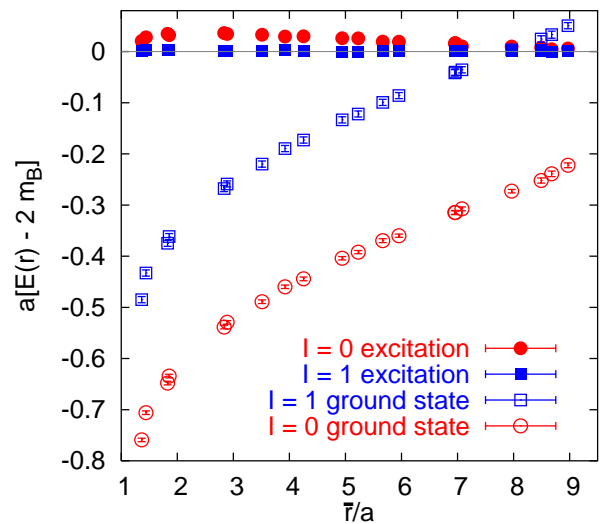}}
\caption {The first two energy levels within the $I=0$ and $I=1$ sectors,
at short distances.}
\label{fig:short1}
\end{figure}

We now focus on the short distance behaviour, both for $I=0$ and for
$I=1$ $B\overline{B}$ systems, and display
the respective two lowest lying
energy levels in Figure~\ref{fig:short1}. 
Note that the difference between the $I=1$ and the $I=0$ ground
states is given by our (unrealistically heavy) $\pi$ mass,
$m_{\pi}\approx 640$~MeV. The lowest state in
the $I=0$ sector corresponds to the conventional
$\overline{Q}Q$ potential. We already noted the bump in the
excited state level, with a maximum
of about 85(10)~MeV, relative to $2m_B$, at a distance of
0.2 -- 0.3~fm. Such a
bump is not present within the $I=1$ sector, to which the exchange
diagram $C_{BB}^{\rm con}$ does not contribute. We assume
this energy barrier to be related to meson exchange. Note that at the
distance of the maximum, $\Delta E(r)\approx 2m_{\pi}$. Unfortunately,
in our study we restricted ourselves to one quark mass
and hence we are unable to investigate the quark mass dependence of
the height and of the position of this feature.

Within  the $I=1$ situation, we also encounter two sectors, namely a
$\overline{Q}Q\pi$ state (with $\overline{Q}Q$ in the
$\Sigma_u^-$ representation that is mass-degenerate
with $\Sigma_g^+$) and the $B\overline{B}$ state
that we label as $|B_a\rangle$. After diagonalization of
this mixing problem one should be able to identify
a mixing angle $\phi$ and the two energy levels\footnote{Note that at
very small distances and/or small
sea quark masses there
will be additional multi-mesonic states between
the $|1_a\rangle$ ground state and the
$|2_a\rangle$ excitation.} $E_1^a$ and $E_2^a$, in analogy to the $I=0$ system.
Again, for $r=0$ the
off-diagonal elements of the corresponding correlation matrix vanish and
the two sectors decouple. The lower lying state will be a
single $\pi$, with
$\overline{Q}$ and $Q$ annihilating.
In contrast, in this limit,
$C_{BB}^{\rm dis}$  will couple to scalar states.
In the colour singlet
sector this will be a scalar $a_0$ meson as well as $\pi\pi$ scattering states.
There will be excitations above these mesonic states,
corresponding to two light quarks, bound to an adjoint static colour source,
in analogy to pure gauge
hybrid potentials where
$\overline{Q}$ and $Q$ do not annihilate at $r=0$
(gluelumps~\cite{Bali:2003jq}).
Note that in the limit $r=0$, the $I=0$ correlation function
$C_{BB}(t)$ will also couple to both,
$B\overline{B}$ states with $q\bar{q}$ in a colour octet (which we shall call
$q\bar{q}$ lumps) and to colour singlet
$f_0/\pi\pi$ states. The latter sector is lighter.

\begin{figure}[th]
\hspace{-0.1in}
\epsfxsize=0.9\columnwidth
\centerline{
\epsffile{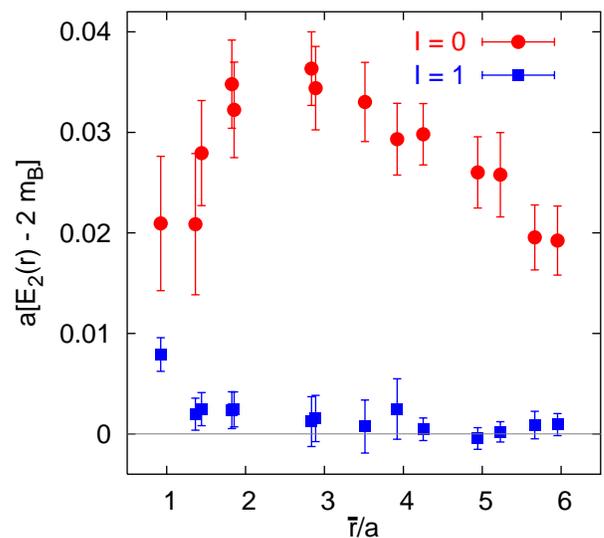}}
\caption {The $I=0$ and $I=1$ excited state energies,
relative to $2m_B$, at short distances, $\overline{r}<r_0\approx 6a$.}
\label{fig:short2}
\end{figure}

As we only have the creation operator
of the $|B_a\rangle$ state ${\mathcal B}_a$ at our disposal
but do not separately investigate the $\overline{Q}Q\pi$ sector
(and the mixing between the two sectors),
we assume the lowest lying $I=1$ state (open squares) to consist
of the ground state potential plus the mass of
the $\pi$.
There will be a (small) correction to this assumption,
due to the interaction energy on a
finite lattice. For the $r>2\sqrt{2}a$ data we are unable to detect this
state within the $C_{BB}^{\rm dis}$ signal, see also Figure~\ref{fig:rat3}.
For $\overline{r}<4a$
we apply two exponential fits with $t_{\min}=2a$.
The overlap with the $\overline{Q}Q\pi$ ground state
turns out to be tiny in these three parameter fits, with mixing angles
ranging from $\sin^2(\phi)=0.0006(22)$ at $\overline{r}\approx 3.51a$
to $\sin^2(\phi)=0.0034(4)$ at $\overline{r}\approx 1.37a$. However,
the fits are consistent with the ground state mass assumption,
$E_1(r)+m_{\pi}$.

In Figure~\ref{fig:short2} we focus on the two $E_2$ levels at
small $r$. As noted before, there is repulsion in the $I=0$ sector
for distances $r>0.25$~fm, with a peak value of the energy barrier of
about 85~MeV. However, at very short range, attraction sets
in. This has to be so since in the $r\rightarrow 0$ limit
the first excitation above
the vacuum is a flavour singlet light quark state, with
$\overline{Q}$ and $Q$ annihilating each other. Hence the form
at asymptotically short distances will be governed by the
perturbative colour singlet potential.

Contrary to Ref.~\cite{Michael:1999nq}, we also observe
(weak) repulsion in the $I=1$ sector. This difference
might be due to a bigger overlap of the ${\mathcal B}_a$
operator used in this previous study with the
$\overline{Q}Q\pi$ ground state.
However, with our operator and statistical accuracy
we are able to clearly separate
the (tiny) $\overline{Q}Q\pi$ pollution from $E_2$.
From a four parameter two exponential fit
to the $I=1$ operator $C_{BB}^{\rm dis}$ at $r=0$
we find, $aE_1=0.394(26)$, very similar to the corresponding
$I=0$ value, $aE_g=0.32(17)$. This might indeed be
a scalar $a_0$ meson. The coupling between our operator and
this state is $a_1^2=0.010(2)$. The first excitation
(with which our operator has 99~\% overlap) that we are able to
resolve is $E_2-2m_B=0.0079(16)a^{-1}$ (left most data point:
$\overline{\mathbf 0}\approx 0.92\,a$). We interpret this as
the lowest lying lump of a $q\bar{q}$ state,
bound to an adjoint static colour
source. In this case the short distance interaction can be
identified with the octet potential~\cite{Bali:2003jq}. As argued above, there
should be further scattering states inbetween the $a_0$ level
and the $q\bar{q}$ lump, however, our operator basis
appears to have (almost) vanishing overlap with them.

\section{Phenomenological implications}
\label{sec:pheno}
We discuss a possible extrapolation of our string breaking
results to the $n_f=2+1$ case with realistic light quark masses.
We also comment on the relevance of the results
with respect to quarkonium spectroscopy.

\subsection{Extrapolation to real QCD}
\label{sec:realqcd}
We expect the string breaking distance to decrease with the
sea quark mass. From the experimental difference $m_{B_s}-m_B=90(3)$~MeV,
we obtain $r_c\approx 2.16\,r_0\approx 1.08$~fm if we assume
invariance of the shape of the $\overline{Q}Q$
potential under variation of the sea quark mass.
This assumption however is rather arbitrary
and we wish to refine this first very rough estimate.

A more controlled way is to extrapolate previous results of the
static-light meson
mass~\cite{Struckmann:2000hm,Bali:2003jv} and of the $\overline{Q}Q$
energy~\cite{Bali:2000vr} quadratically in the $\pi$ mass. The latter
extrapolation has already been performed in Ref.~\cite{Bali:2000vr}.
For the static-light mass
we obtain an upward shift,
$\Delta m_B=0.21(10)\,r_0^{-1}$, when
replacing our simulated quark mass by the physical light quark mass.
This direction of change
is possible since the self energy of the static propagator
increases with decreasing sea quark mass~\cite{Bali:2002wf}.
The potential at $r_c$ also moves upwards, unsurprisingly by an amount
that is larger than $2\Delta m_B$.
In combining the two extrapolations we obtain
$r_c=(2.27\pm 0.20)\,r_0\approx (1.13\pm 0.10)$~fm
for $n_f=2$ light sea quarks,
in good agreement with the rough phenomenological estimate presented above.

The previous lattice results~\cite{Bali:2000vr,Struckmann:2000hm,Bali:2003jv}
were obtained with a static action that differs from
the present one where we employ fat temporal links
(see Sec.~\ref{sec:statq}), however, this change will not affect the string
breaking distance since, when introducing the fat link action,
both energy levels
are always shifted downwards by the same amount.

We discuss
the effect of a third, heavier, sea quark flavour.
In this case there will be two separate thresholds, one for the decay into
what we call $B$ and $\overline{B}$ mesons and one into
$B_s$ and $\overline{B}_s$.
It is not {\em a priori} clear what effect the inclusion of such
a third sea quark
has on the $r_c$ position at which the decay into
$B\overline{B}$ sets in.
A comparison between the $n_f=0$ and the $n_f=2$ situations might give some
indication. Interpolating the $n_f=0$ static-light masses of
Refs.~\cite{Allton:1994jz,Ewing:1995ih} to our quark mass,
$m_{\pi}/m_V=0.704(5)$, we obtain the value
$m_B=0.540(10)\,a^{-1}$ at $\beta=6.2$ where $r_0=7.30(4)\,a$.
Together with the potential
from Ref.~\cite{Bali:1992ru}, this corresponds to $r_c=2.53(8)\,r_0$,
very consistent with our $n_f=2$ result, Eq.~(\ref{eq:RC}),
$r_c=2.50(3)\,r_0$. So we would expect
the value,
\begin{equation}
r_c=(2.27\pm 0.20)\,r_0\approx (1.13\pm 0.10)\,\mbox{fm},
\end{equation}
to remain largely unaffected
by the addition of the strange quark.
Note that there are
additional systematic errors of about $5~\%$ on
the scale $r_0$ and that we have not attempted
a continuum limit extrapolation. We expect
large distance physics like the string breaking scenario
to remain largely unaffected by charm quark dynamics
which, however, might influence short distance
interactions.

\begin{figure}[th]
\hspace{-0.1in}
\epsfxsize=0.9\columnwidth
\centerline{
\epsffile{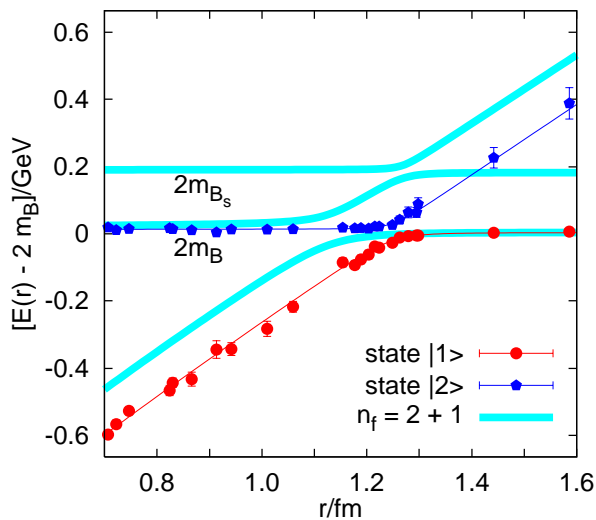}}
\caption {The energy levels in physical units for $n_f=2$ at a quark mass
of slightly less than the strange quark mass (data points). The bands
represent our $n_f=2+1$ speculation.}
\label{fig:final}
\end{figure}

In Figure~\ref{fig:final} we display our $n_f=2, m\lesssim m_s$ energy levels
in physical units. The plotted parametrizations are,
\begin{eqnarray}
E_1(r)&=&2m_B+g_1(r)V(r)+c_1,\\
E_2(r)&=&2m_B+[1-g_2(r)]V(r)+c_2,
\end{eqnarray}
where
\begin{eqnarray}
g_i(r)&=&\frac{1}{2}-\frac{1}{\pi}\arctan\left[d_i(r-r_c)\right],\\
V(r)&=&-e\left(\frac{1}{r}-\frac{1}{r_c}\right)+\sigma \left(r-r_c\right).
\end{eqnarray}
We use the $\arctan$ function in the definition of
the smeared out step functions $g_i(r)$, rather than e.g.\ $\tanh$,
to allow for a direct comparison
with the dependence of the mixing angle $\theta$ on $r$,
Eq.~(\ref{eq:thetafit}). Also note that the above parametrizations 
represent
only effective descriptions of the data, within a certain window
of distances $r<1.6$~fm. For instance, $E_1$ does not
have the correct large distance limit $2m_B$.
The parametrization of $E_1(r)$ is valid for $r>0.2$~fm, while that
of $E_2(r)$ applies to $r>0.75$~fm. In this latter channel,
we encounter
a repulsive potential barrier at smaller distances, see
Figures~\ref{fig:break1}, \ref{fig:short1} and \ref{fig:short2}.

We use the same $e\approx 0.36$, $\sqrt{\sigma}\approx 447$~MeV as in
Eqs.~(\ref{eq:funnel2}) -- (\ref{eq:funnel3}) and the
$r_c\approx 1.25$~fm of Eq.~(\ref{eq:RC}). We obtain
$c_1\approx -21$~MeV and $c_2\approx 31$~MeV. Note that
$\Delta E_c=c_2-c_1\approx 51$~MeV. The parameters $d_i$ read,
$d_1\approx 2.7\,\mbox{GeV}\approx (0.073\,\mbox{fm})^{-1}$,
and $d_2\approx 4.0\,\mbox{GeV}\approx (0.049\,\mbox{fm})^{-1}$.
Note that we parameterized the mixing angle in the string breaking
region in a similar fashion, Eq.~(\ref{eq:thetafit}), with
$d\approx 5.5$~GeV.

We speculate about the real QCD situation in Figure~\ref{fig:final}
(bands).
Besides  the above discussed reduction of the
string breaking distance, we would
expect the shape of the energy gap $\Delta E(r)$ to depend
on the quark mass as well. A lighter mass will result in a larger
gap $\Delta E_c$ and a broadened mixing region. We plot the corresponding
curves with the arbitrary correction factors,
$c_i\mapsto 1.5\,c_i$, $d_i\mapsto d_i/1.5$, also taking
into account an increase of $e\approx 0.4$ and
a reduction of $\sqrt{\sigma}\approx 440$~MeV~\cite{Bali:1998pi}.
There will be a second level crossing around $2m_{B_s}$, which we
also sketch in the Figure.

Another uncertainty is related
to the short distance dynamics that we observe.
We found an 85~MeV high potential barrier
within $E_2(r)$, at a distance of about 0.2~fm. Most likely this is related to
$\pi$ exchange~\cite{Barnes:1999hs,Tornqvist:2004qy}. In this case the dimensions, both
of the height of the barrier and of its position should be provided
by the $\pi$ mass $m_{\pi}$. On one hand,
reducing $m_{\pi}$ by a factor of {\em four}
down to its physical value could easily
move this region close to the string breaking distance. On the other hand,
we would then expect the associated correction to the $E_2(r)$ level
around $r_c$ to be smaller than 20~MeV, while $\Delta E_c>50$~MeV. 

Finally, we remark that at smaller quark masses
additional scattering states will occur between
the $E_1(r)$ and the $E_2(r)$ energy
levels, at short distances where $\Delta E(r)>2m_{\pi}$.
It is clear that replacing our qualitative
$n_f=2+1$ picture by a truly quantitative understanding would require
simulations at
additional light quark mass parameters. 

\subsection{Relation to quarkonium physics}
String breaking provides an intuitive example of a strong decay.
In addition, static potentials can readily be related to
quarkonium physics: this can be achieved by introducing a phenomenological
Born-Oppenheimer approximation~\cite{Eichten:1978tg} or, more
systematically, within the framework of
potential NRQCD (pNRQCD)~\cite{Brambilla:2004jw}.
As long as the quarkonium state in question is much lighter than
the respective strong decay threshold into a pair of heavy-light
mesons, the ground state energy level $E_1(r)$ should, to leading
order in the relative $\overline{Q}Q$ velocity $v\ll 1$,
accurately
encapsulate the relevant physics.
If this is not the case anymore,
then additional terms have to be added to the pNRQCD Lagrangian,
to incorporate the $\overline{Q}q\bar{q}Q$
sector and transitions between the two sectors:
when it comes to strong decays like $\Upsilon(4S)\rightarrow
B\overline{B}$,  $E_2(r)$ and
the transition rate/coupling $g(r)$ [or, equivalently, the mixing angle
$\theta(r)$] are required to
describe the system, in addition to $E_1(r)$.

For states that are stable against such decay, but
which are in the vicinity of a threshold,
mixing effects will result
in mass shifts~\cite{Eichten:1975ag,Eichten:1978tg}.
This has been discussed in some
detail for instance in
Refs.~\cite{Eichten:2004uh,Barnes:2004fs,Tornqvist:2004qy} for the recently
discovered $X(3872)$ charmonium state~\cite{Choi:2003ue,Brambilla:2004wf}.
In this context, our results could hint towards the
nature of the underlying
interaction Hamiltonian and of the $q\bar{q}$
pair creation mechanism that is at work.

We have demonstrated in Sec.~\ref{sec:transition}
that mixing is sufficiently weak to allow
a basis of $\overline{Q}Q$ and $\overline{Q}q\bar{q}Q$ quark model
Fock states to be 
a valid starting point in any such analysis~\cite{Swanson:2005rc}.
However, not only in
the string breaking region but also at short distances the
ratio $g(r)/\Delta E(r)$ can be sizeable. We find $g(r)$ to be of
an $O(100\,\mbox{MeV})$ magnitude which is typical for
strong decay dynamics. $g(r)$ sets out from {\em zero} at the origin,
increases
to about 320 MeV around $r\approx 0.2$~fm and
reduces to $\Delta E_c/2 \approx 25$ MeV,
in the string breaking
region. The maximum value is due to meson exchange and its position 
coincides with, $\Delta E(r)\approx 2 m_{\pi}\approx 4g(r)$.
We would expect this medium range $g$ value to somewhat
decrease with lighter quark masses and
$g(r_c)$ to increase.

In QCD with sea quarks there are not only $B\overline{B}$ excitations present
but also $\overline{Q}Q$-gluon hybrid potentials. These however are
energetically higher and, unless we are interested in hybrid quarkonia
with spin-exotic quantum numbers, not a dominant
correction~\cite{Burch:2003zf} to
quark potential model predictions. Obviously, hybrid meson mixing
and decay is interesting in itself~\cite{McNeile:2002az}, and, in this context,
a detailed study of the breaking of hybrid strings would be interesting.
Finally, at light sea quark masses
hadronic transitions between quarkonia, mediated by $\pi$ radiation,
become possible, the inclusion of which necessitates further
modifications.
\section{Summary}
We were able to resolve the string breaking problem in $n_f=2$ QCD,
at one value of the lattice spacing $a^{-1}\approx 2.37$~GeV and
of the sea quark mass, $m\lesssim m_s$.

To achieve this result, the systematic improvement of our methods
beyond the latest lattice technology was crucial.
In particular, we used highly optimized smearing
functions to enhance the overlap of our test wave functions with the
physical states, we employed an improved static action
and we realized many off-axis source separations.
We used stochastic estimator techniques
(SET) to calculate all-to-all propagators. The variance of SET was reduced
by exactly calculating the contribution from the lowest lying eigenmodes
of $\gamma_5M$, where $M$ denotes the Wilson Dirac matrix.
Further variance reduction was achieved by the (new) hopping parameter
acceleration (HPA) technique. These methods, most of which are
neither specific to the string breaking problem nor to the static charge sector
of lattice QCD, can readily be applied to
a large spectrum of problems. In general we would expect the gain factor from
HPA to decrease at
lighter sea quark masses while the improved convergence of TEA should
compensate for this.

We determined a mixing angle $\theta(r)$. Truncating the Fock basis
after states containing four quark operators, 
the $\overline{Q}Q$ component of the
physical ground state was given by
$\sin\theta$ and the $B\overline{B}$ content
by $\cos\theta$.
We distinguished between the explicit and the implicit detection of
mixing effects: a non-vanishing transition element between
$\overline{Q}Q$ and $B\overline{B}$ states is an explicit signal of
mixing. Additionally, at $r<r_c$ the
ground state energy will dominate, even within the
$B\overline{B}$ operator, at large Euclidean times (implicit mixing).
We were able to verify this behaviour.
At large $r>r_c$, the lowest lying state will have a mass slightly smaller
than twice the static-light mass, $2m_B$, and dominantly
couple to the $B\overline{B}$ operator. We were unable to detect this
signal in the $\overline{Q}Q$ sector alone.
Based on our mixing analysis, we expect such an implicit detection
of string breaking to be almost hopeless, as
high precision
Wilson loop data at Euclidean times $t=O(5~\mbox{fm})$ would be necessary
--- and this at distances $r>1.2$~fm!

We defined two string breaking distances: $r_c=15.00(8)a\approx 1.248(13)$~fm
denotes the distance at which the energy gap between the two levels
$\Delta E(r)=E_2(r)-E_1(r)$ assumes its minimal value $\Delta E_c=0.022(1)a^{-1}
\approx
51(3)$~MeV while $r_s=14.95(12)a\approx 1.244(16)$~fm denotes the distance
of perfect mixing between the two states,
in terms of the mixing angle $\theta(r_s)=\pi/4$. 
The conversion into physical units has been made by setting $r_0=0.5$~fm.
Note that
$r_c-r_s=0.053(53)a$ is compatible with zero, within a standard deviation
of less than $5\times 10^{-3}$~fm.

We would expect $\Delta E_c\approx 51(3)$~MeV to
increase with lighter sea quark masses and hence this value
should be regarded as a lower limit to the case with massless or very light
sea quarks. In real QCD also the string breaking distance should decrease.
We estimate $r_c= 1.13(10)(10)$~fm for this case.
The first error is statistical
and from the chiral extrapolation of previous results, the second
incorporates possible finite lattice spacing effects, the scale uncertainty in
$r_0=0.5$~fm and effects due to the inclusion of a third active
sea quark flavour. The qualitative situation is depicted in
Figure~\ref{fig:final}.

We can define a transition rate $g(r)=\Delta E(r)\sin[2\theta(r)]/2$
between $\overline{Q}Q$ and $B\overline{B}$
states and note that $g(r_s)=\Delta E(r_s)/2$ and hence
$g(r_c)\approx \Delta E_c/2$. In the large $N_c$ limit
this means that at leading order, $\Delta E_c\propto\sqrt{n_f/N_c}$,
if we are interested in the screening of a fundamental string
by a sea of $n_f$ massless flavours of fundamental particles,
e.g.\ scalars or quarks. For the breaking of an adjoint string into
two gluelumps the expectation reads, $\Delta E_c\propto 1/N_c$.
In view of the precision of the $n_f=2$ QCD results presented here,
it should be worthwhile to dedicate renewed effort onto
string breaking studies of $SU(N_c)$ gauge theories with and
without Higgs fields, to confirm the expectations, and to explore the
applicability of large $N_c$ arguments to strong hadronic decays.

We conclude that our study constitutes an important step towards the
understanding of mixing effects and strong decays in
quarkonium systems~\cite{Eichten:2004uh,Swanson:2005rc}.
Studying the energy between pairs of
static-light mesons can also be viewed as
a milestone with respect to
a future calculation of $\Lambda_Q\Lambda_Q$ forces,
which are related to nucleon-nucleon interactions~\cite{Arndt:2003vx}.
\begin{acknowledgments}
We thank Philippe de Forcrand, Jimmy Juge, Owe Philipsen and the
participants of the third Quarkonium Working Group Workshop at IHEP Beijing
for discussions and Sara
Collins, Elvira Gamiz, Zdravko Prkacin, Eric Swanson
and Hartmut Wittig for comments.
We thank the John von Neumann Institute for Computing (NIC) for granting
us computer time. The computations have been performed on the IBM Regatta
p690+ (Jump) of ZAM at FZ-J\"ulich and on the ALiCE cluster computer
of Wuppertal University. We thank the staff at ZAM for
their support, in particular Norbert Attig.
This work is supported by the
EC Hadron Physics I3 Contract No.\ RII3-CT-2004-506078.
The development of the parallel file system
for ALiCE was supported by the Deutsche Forschungsgemeinschaft under grant Li
701/3-1. G.B.\ is supported by PPARC. 
Th.D.\ is supported by the DFG under
grant Li 701/4-1.
\end{acknowledgments}


\end{document}